\documentclass[iop]{emulateapj}

\usepackage{lineno}  
\usepackage{amsmath} 
\usepackage{natbib}
\setcitestyle{round,aysep={},yysep={;}}

\usepackage{graphicx}
\usepackage{color}
\newcommand{\be}{\begin{equation}}
\newcommand{\ee}{\end{equation}}

\shorttitle{MW dark matter/MOND from vertical stellar kinematics}
\shortauthors{L\'opez-Corredoira}

\begin{document}

\title{Milky Way dark matter distribution or MOND test from vertical stellar kinematics with Gaia DR3}
\author{Mart\'\i n L\'opez-Corredoira$^{1,2,3}$}
\altaffiltext{1}{Instituto de Astrof\'\i sica de Canarias, E-38205 La Laguna, 
Tenerife, Spain; martin@lopez-corredoira.com}
\altaffiltext{2}{PIFI-Visiting Scientist 2023 of Chinese Academy of Sciences at Purple Mountain Observatory, Nanjing 210023,\\ and National Astronomical Observatories, Beijing 100012}
\altaffiltext{3}{Departamento de Astrofisica, Universidad de La Laguna, E-38206 
La Laguna, Tenerife, Spain}

\begin{abstract}
Vertical stellar kinematics+density 
  can be used to trace the dark matter distribution [or the equivalent phantom mass
in a Modified Newtonian Dynamics (MOND) scenario] through Jeans equations. 
In this paper, we want to improve this type of analysis by making 
use of the recent data of the 6D information from the Gaia-DR3 survey in the
anticenter and the Galactic poles to obtain the dynamical mass distribution near plane regions, including extended kinematics over a wide region of 8 kpc$<R<$22 kpc, $|z|<3$ kpc.

Our conclusions are as follows: (i) the model of the spherical dark matter halos and the MOND model are 
compatible with the data; (ii) the 
model of the disky dark matter (with density proportional to the gas density) is excluded; (iii) the
total lack of dark matter
(there is only visible matter) within Newtonian gravity is compatible with the data;
for instance, at solar Galactocentric radius, we obtained 
$\Sigma=39\pm 18$ M$_\odot$ pc$^{-2}$ for $z=1.05$ kpc, compatible with the expected value
for visible matter alone of 44 M$_\odot$ pc$^{-2}$, thus allowing zero dark matter.
Similarly, for $R>R_\odot$, $z=1.05$ kpc: $\Sigma=28.7\pm 9.6$, $23.0\pm 5.7$, $16.9\pm 5.8$, 
$11.4\pm 6.6$ M$_\odot$ pc$^{-2}$, respectively, for $R=10,13,16,19$ kpc, 
compatible with visible matter alone.
Larger error bars in comparison with previous
works are not due to worse data or a more awkward technique but to a stricter modeling of the stellar distribution.
\end{abstract}

%\keywords{Galaxy: kinematics and dynamics -- dark matter -- Galaxy: halo -- Galaxy: disk}

%________________________________________________________________

\section{Introduction}

The hypothesis of the existence of dark matter halos in galaxies like the Milky
have different motivations, either from cosmology, extragalactic astronomy,
or investigations in our own Galaxy \citep[e.g.,][\S 3.3]{Lop22}.

One of the first observational evidence for the missing mass in our Galaxy was provided
 by the fact that the Galactic mass derived from the
motions of distant globular clusters was approximately three times larger than
that obtained from the rotation of the inner disk of the Galaxy \citep{Fin63}.
Over many decades, the rotation curve of the Galactic disk in the gas component \citep[e.g.,][]{Cle85,Bra93} 
or the stellar component \citep[e.g.,][]{Pon97,Bat13} or masers \citep[e.g.,][]{Hon12,Rei14}
has also been investigated, or the motion of satellite galaxies orbiting the Milky Way \citep[e.g.,][]{Sof15,Cal19}.
However, velocities in galaxy pairs and satellites might also measure the mass of the intergalactic 
medium filling the space between the members of the pairs rather than the mass of dark halos 
supposedly associated with the galaxies \citep{Lop99,Lop02b}. 

In spite of the large number of studies, the consensus knowledge of the characteristics of the dark matter
halo derived from rotation curves is still small. 
First, there is a huge dispersion of the values of the total mass of the Galaxy, including this halo,
with values as small as $3-4\times 10^{11}$ M$_\odot $ \citep{Sof09,Ou24}, or as large as 
$2.2\times 10^{12}$ M$_\odot $ \citep{Sak03}. Second, there is not even consensus on whether the rotation
curve is flat \citep{Oll00,Sof09b} or Keplerian \citep{Hon96,Gal15,Jia23} over some Galactocentric distance in the outer 
disk or something in between with a slight decline \citep{Koo24}.

Rotation curves in spiral galaxies can also be explained with hypotheses different from 
the standard dark matter halos of non-baryonic cold dark matter \citep[Section 3.6]{Lop22}: with different types of dark matter, even baryonic, magnetic fields, non-circular orbits in the outer disk, alternative gravity theories.
The most popular alternative to dark matter is the modification of gravity laws proposed in Modified Newtonian
Dynamics (MOND; \citet{San02,Fam12}), which modifies the Newtonian law for accelerations lower than a constant $a_0$,
 or dark matter distributed in the outer disK instead of a halo  \citep{Bos81,Pfe94,Fen15,McK15,Fer16,Kra16,Sip21,Syl23,Syl24,Syl24b} (disky dark matter).
Here, we will focus on MOND and disky dark matter apart from the standard (spheroidal or slightly ellipsoidal) halo dark matter.

Apart from rotations curves, there are other ways to quantify or trace the putative dark matter distribution, although
can indeed be solved without dark matter: galactic stability \citep{Too81} 
or warp creation \citep{Lop02b}, for instance. 

Kinematics of stars can be used in a different way to trace the dark matter distribution (or the equivalent phantom mass
in a MOND scenario), particularly using the vertical velocities of stars rather than the rotation speed.
\citet{Mon12} used Jeans equations to relate dispersion of vertical stellar velocities, stellar density
distribution, and dark matter density. The dynamical surface mass density at the solar position with a distance to the Galactic plane between $z=1.5$ and 4 kpc was estimated using the velocity dispersion of thick disk stars \citep{Mon12a}, resulting 
in compatibility with the expectations of visible mass alone. No dark component was required to account for the 
observations, all the current models of a spherical dark matter halo were excluded at a confidence level higher than $4\sigma $. 
This result has been criticized on several grounds \citep[e.g.,][]{San12,Bov12}, such as the calculation of the velocity 
dispersion or the assumptions in the calculations. The criticisms have been responded to \citep{Mon15}.
Indeed, the general impression among many different authors is that the measured mass density is highly dependent on the assumptions using the Jeans equations \citep[e.g.,][]{Can16,Che24}. 
In particular, the consideration of the disk as a double
component of thin+think disks, the vertical profile of stellar density (exponential or sech$^2$), the flare, and other considerations matter \citep{San16,Nit21}.

In this paper, we want to improve upon this type of analysis by revising the equation, taking into account subtle
details of the distribution in Section \ref{.dens.kin},  making 
use of the recent data on the 6D information from the Gaia-DR3 survey \citep{Gai23}, 
including extended kinematics over a wide region up to $R>20$ kpc by means of a deconvolution technique of the parallax 
errors in
Section \ref{.GaiaDR3}.
We will also use accurate information on the stellar and gas+dust density distribution, considering their error bars.
All of this will serve to derive the dark (total - visible) matter distribution in the range $8<R({\rm kpc})<22$, $|z|<3$ kpc, which may distinguish between spheroidal halo dark matter and disky dark matter in Section
\ref{.dark}.
We also carry out the full analysis under the considerations of a MOND scenario in Section \ref{.MOND}. 
Systematic errors are considered in Section \ref{.syst}. A 
discussion is given in Section \ref{.concl}.

\section{Density derived from stellar kinematics}
\label{.dens.kin}

Given an axisymmetric mass distribution $\rho (R,z)$, the surface density within a finite distance $z$ of the plane is defined as
\begin{equation}
\label{surf.dens}
\Sigma (R,z)\equiv \int _{-z}^z\rho (R,z') \mathrm{d}z'
.\end{equation}
Using Poisson equation, we get 
\begin{equation}
4\pi G\Sigma(R,z)=\frac{1}{R}\int_{-z}^{z} \frac{\partial [R\,a_R(R,z')]}{\partial R} \mathrm{d}z'
\end{equation}\[
+[a_z(R,z)-a_z(R,-z)]
,\]
where $a_R$ (related to the rotation curve as $V_c^2=R\,a_R$) and $a_z$ are the accelerations due to the given potential 
and, assuming steady equilibrium, can be related through the Jeans equations
with the kinematics of the stellar distribution with density of $\rho _*(R,z)$ \citep{Mon12,Loe12,Vil22}:
\begin{equation}
V_c^2(R,z)=R\,a_R(R,z)=-\frac{R}{\rho _*(R,z)}\frac{\partial [\rho_*(R,z)\overline{V_R^2}(R,z)]}{\partial R}
\end{equation}\[
-\frac{R}{\rho _*(R,z)}\frac{\partial [\rho_*(R,z)\overline{V_RV_Z}(R,z)]}{\partial z}-\overline{V_R^2}(R,z)+\overline{V_\phi^2}(R,z)
,\]
\begin{equation}
\label{az}
a_z(R,z)=-\frac{\overline{V_RV_z}(R,z)}{R}-\frac{1}{\rho _*(R,z)}\frac{\partial [\rho_*(R,z)\overline{V_RV_z}(R,z)]}{\partial R}
\end{equation}\[
-\frac{1}{\rho _*(R,z)}\frac{\partial [\rho_*(R,z)\overline{V_z^2}(R,z)]}{\partial z}
.\]

Note that in the Jeans equations, we use the stellar density $\rho _*$ and not the total density $\rho $ because our distribution function
over which we measure the velocities is the stellar population alone; we have no information about the velocities of dark matter particles. 
This does not mean that we are neglecting the dark matter and gas here, but
we separate the stellar component for our distribution function, embedded in a potential that includes everything (stars, gas, and dark matter). 
There is no problem of lack of self-consistency here.
The Jeans equations describe the motion of any collection of particles in a gravitational field (e.g., \citet[Section 4.8]{Bin08},\citet[ch. 10]{Cio21},\citet{Mon12,Bov12}). 
For a given gravitational field, each component of matter obeys a separate collisionless Boltzmann equation from which the Jeans equations are derived. 
In fact, even the distribution for any particular subtype of stars 
would obey a separate equation if we could measure the density and velocities of this subcomponent. 
In our case, we use a multicomponent with the density and velocity distribution average of all of the
observed single subtypes of stars, where the Jeans equations keep the same dependence (see
Appendix \ref{.Jeansmulti}).

The rotation curve is almost flat, with a very small decrease with $R$ \citep{Wan23} ($\beta $). We assume:
\begin{equation}
V_c(R,z)^2=[V_c(R_\odot ,z)+\beta (z)(R-R_\odot )]^2
\end{equation}\[
\approx V_c^2(R_\odot ,z)+2\beta (z)V_c(R_\odot ,z)(R-R_\odot )
,\]
and $V_z^2(R,z)=-V_z^2(R,-z)$.
Also, the terms with $\overline{V_RV_Z}$ in Eq. (\ref{az}) and its gradients can be neglected \citep{Can16,Eil19}, because they are around two orders of magnitude smaller compared to the remaining terms (see also \S \ref{.rms} including off-plane regions).

The stellar density is modeled with a thin+thick disk, with $f$
 the ratio of thick-to-total disk stars in the solar neighborhood,
 radial exponential dependence of scale length $h_R$, and vertical
exponential or sech$^2$ dependence with scale height $h_z$ of each of the thin and thick subcomponents. 
There is plenty of literature discussing whether exponential of sech$^2$ functions fit better
the vertical distribution \citep[e.g.,][]{Dob20,Eve22,Chr22,Vie23}, but 
in any case, both forms are almost equivalent at large $|z|$.
The contribution of halo stars in the regions used, $|z|<3$ kpc, is negligible \citep{Bil08}. We
include a flare (that is, variable scale heights instead of constant; for a more general expression of Jeans equations with
flare, see also \citet[Eq. 13]{Lop20} or \citet{Nit21}):

\begin{enumerate}

\item Vertical distribution as the sum of two exponentials:
\begin{equation}
\label{rhoestexp}
\rho _*(R,z)=\rho _{*,\odot}\exp\left(-\frac{R-R_\odot}{h_R}\right)
\end{equation}\[
\times \left[(1-f)\exp\left(-\frac{|z|}{h_{z,{\rm thin}}(R)}\right)
+f\exp\left(-\frac{|z|}{h_{z,{\rm thick}}(R)}\right)\right]
,\]

\item or vertical distribution as the sum of two sech$^2$ functions:

\begin{equation}
\label{rhoest}
\rho _*(R,z)=\rho _{*,\odot}\exp\left(-\frac{R-R_\odot}{h_R}\right)
\end{equation}\[
\times \left[(1-f){\rm sech}^2\left(\frac{|z|}{2h_{z,{\rm thin}}(R)}\right)
+f\,{\rm sech}^2\left(\frac{|z|}{2h_{z,{\rm thick}}(R)}\right)\right]
,\]

\end{enumerate}

With all of these ingredients, we get
\begin{equation}
\label{4pigro}
4\pi G\Sigma(R,z)\approx \frac{2V_c(R_\odot ,z)\int_{-z}^{z} \beta (z')\mathrm{d}z'}{R}
-2\frac{\partial \overline{V_z^2}(R,z')}{\partial z}
\end{equation}\[
+\frac{2\overline{V_z^2}(R,z)}{\overline{h_z}(R,z)}
,\]

\begin{enumerate}
\item In a vertical distribution as the sum of two exponentials (Eq. \ref{rhoestexp})
\begin{equation}
\label{hzexp}
\overline{h_z}(R,z)=
\end{equation}\[
\left[\frac{\frac{(1-f)}{h_{z,{\rm thin}}(R)}+\frac{f}{h_{z,{\rm thick}}(R)}\exp\left(|z|\frac{h_{z,{\rm thick}}(R)-h_{z,{\rm thin}}(R)}{h_{z,{\rm thin}}(R)h_{z,{\rm thick}}(R)}\right)}{(1-f)+f\exp\left(|z|\frac{h_{z,{\rm thick}}(R)-h_{z,{\rm thin}}(R)}{h_{z,{\rm thin}}(R)h_{z,{\rm thick}}(R)}\right)}\right]^{-1}
\]

\item In a vertical distribution as the sum of two sech$^2$ functions (Eq. \ref{rhoest})
\[
\overline{h_z}(R,z)=
\]\[
\left[\frac{\frac{(1-f)}{h_{z,{\rm thin}}(R)}\tanh (x_{\rm thin})+
\frac{f}{h_{z,{\rm thick}}(R)}\frac{{\rm sech}^2(x_{\rm thick})}{{\rm sech}^2(x_{\rm thin})}
\tanh (x_{\rm thick}) } 
{(1-f)+f  \frac{{\rm sech}^2(x_{\rm thick})}{{\rm sech}^2(x_{\rm thin})} }
\right]^{-1}
,\]\begin{equation}
\label{hz}
x_{\rm thin}=\frac{|z|}{2h_{z,{\rm thin}}(R)};\ x_{\rm thick}=\frac{|z|}{2h_{z,{\rm thick}}(R)}
\end{equation}
\end{enumerate}

Here, $\overline{h_z}(R,z)$ stands for an equivalent average global scale height (or scale at which the variation of the density changes by factor $e$
in the vertical direction),
including thin and thick disks and their flares.
The expression using the vertical distribution as a sum of two exponential 
has an inconsistency at $z=0$, where we get $\Sigma (z=0)>0$ if $\overline{V_z^2}(R,z=0)\ne 0$, as it is the real case, whereas it should be zero
according to our own definition in Eq. (\ref{surf.dens}).
This derives from using a sum of exponential stellar distributions, which has $\frac{1}{\rho _*}\frac{\partial \rho _*(z\rightarrow 0)}{\partial z}=\frac{1}{h_z}$ 
for each of the thin and thick components. However, if we assume in the limit of low $z$ at $\rho _*\propto {\rm sech}^2(|z|/2h_z)$, we
get instead $\frac{1}{\rho _*}\frac{\partial \rho _*(z\rightarrow 0)}{\partial z}=\frac{\tanh[|z|/(2h_z)]}{h_z}\approx \frac{|z|}{2h_z^2}$ for low $|z|$, converging
to zero for the limit of $z=0$.
For this reason, we will adopt in the application throughout this paper 
the vertical distribution as the sum of two sech$^2$ functions, although some application of the exponential profile will be carried out to check for consistency or differences.
In any case, at intermediate-high $z$, both expressions are almost equivalent.

In practice, we can determine $\overline{V_z^2}(R,z)=\overline{V_z}^2(R,z)+\sigma _{V_z}^2(R,z)$. The
average $\overline{(...)}(R,z)$ includes bins with distance from the plane $z$.  
Both the average and dispersion of vertical velocities are provided from kinematic distribution of
stars. Most authors \citep[e.g.,][]{Mon12,Vil22,Zhu23} assume that $\sigma _{V_z}$ falls exponentially
with $R$, but this is not correct in a flared disk, and as a matter of fact, it is observed that
$\sigma _{V_z}$ has a minimum around $R=15$ kpc and it grows for larger $R$ \citep{Wan23}.
We assume $R_\odot =8.34$ kpc throughout this paper.

\section{Analysis from extended kinematic maps of Gaia DR3}
\label{.GaiaDR3}

\subsection{Extended kinematic maps}
\label{.extended}

We derive kinematic from the subsample of Gaia DR3, including 
radial velocities,\footnote{For a total of 32 million stars \citep{Kat23}, Gaia RV targets were selected to include sources with $G_{RVS}<14$ covering the whole range of temperatures of the old population between F4 and M7 spectral types of any luminosity class.} 
and carry out a deconvolution of parallax errors through an inversion technique based on Lucy's method \citep{Lop19}, allowing to reach much larger distances than with the 
direct determination of distance as the inverse of parallax for each star. The methodology and application
to Gaia data have already been explained in previous works \citep{Lop19,Wan23}. 
We have not included zero-point correction in the parallaxes of Gaia data. This has a negligible effect 
on the kinematic maps within the
explored range of distances \citep{Lop19,Wan23}.

Here, we repeat the analysis of \citet{Wan23},
adding a couple of improvements: (1) we remove the high-velocity stars ($|V_R|>100$ km/s, 
$|V_z|>100$ km/s, $V_\phi <0$, $V_\phi >400$ km/s), possibly related to the nonequilibrium state of the Galactic disk, and calculate the dispersion of velocities ($\sigma _V$) taking into account this cut in the tail of the Gaussian distribution (see Appendix \ref{.nonhighvz}); (2) we calculate $\sigma _V$ directly from the dispersion of the velocity of stars once we know the
average distance of each bin with a given parallax from Lucy's method, rather than an expansion as done in \citet{Wan23}. \footnote{\citet{Wan23}'s method: $\sigma _{V_z}=\sqrt{\sin ^2(b) [\sigma_{V_r}^2 - \Delta {V_r} ^2] + \cos^2(b)\, r^2[\sigma_{\mu _b}^2 - 
\Delta {\mu _b}^2] }$, where $r$ is the heliocentric distance; $\sigma_{V_r}$  and $\Delta V_r$
are, respectively, the dispersion and error in radial velocities; $\sigma_{\mu_b}$ and $\Delta {\mu_b}$
are, respectively, the dispersion and error in proper motions 
in the direction perpendicular to the Galactic plane.} Our
new calculation is more accurate, whereas \citet{Wan23}'s method gives slightly different results possibly due to non-negligible
covariance terms. 
Our results for the median velocities, $\overline{V_z}(R,z)$, are in agreement with those obtained by \citet{Wan23}, 
and the $\sigma _{V_Z}$ are slightly lower, especially in off-plane regions because 
\citet{Wan23} included few high-velocity stars that inflate $\sigma _{V_Z}$, while the results here are more
reliable.

Here, we only use data in two directions:
\begin{description}

\item[Galactic poles] $|b|>60^\circ $, $\forall \ell$, $R_\odot -1<R<R_\odot +1$ kpc. This reaches the highest distance from the plane in Galactocentric
radii close to the solar one. 

\item[Anticenter] $160^\circ <\ell <200^\circ $, $\forall b$. This allows it to reach a higher distance in 
the outer disk.

\end{description}
These lines of sight are little affected by extinction, so we can go farther in distance, and we are
less affected by the uncertainties in the determination of average distance through the deconvolution technique.

\subsection{Warp}

The density distributions are not affected by the warp in these lines of sight of the anticenter because the amplitude of the
warp is almost null in the azimuth corresponding to the anticenter (whereas the maximum absolute amplitude corresponds to
azimuths around 80$^\circ $ and 260$^\circ $) \citep{Chr22}.

The average vertical velocity is slightly changed by the warp \citep{Lop20}, but given the symmetry between northern and southern Galactic hemispheres, on average for both the positive and the negative Galactic latitudes, the effect of the warp within distances $z$ from the plane is null.
The dispersion of velocities, which is the relevant term used in the Jeans equations, is not affected by the warp.

Some researchers have speculated that the outer disk is in a nonstationary state due to some precession of the warp \citep[e.g.,][]{Pog20}. 
However, doubt has been cast upon the precession of the warp \citep{Chr21}. In any case, even in the case of significant warp precession, it would affect
few km/s (negligible) in the average vertical velocities toward the anticenter, and it does not affect the dispersion of velocities.

\subsection{Scale height and flare of the thin and thick disks}

Calculating the scale heights directly from
the data of stars only with radial velocities has many uncertainties 
because this sample is incomplete, the luminosity function of the population is difficult
to model, and there are some selection functions that need some corrections, which would make the calculation quite inaccurate. Instead, we will use the measurements of the scale heights for a complete sample with $G<19$ in Gaia 
data, which only needs the luminosity function to derive the distribution of all stars.

\citet{Chr22} derived the average scale height of the 'whole population' of 
Gaia EDR3/$G<19$+Lucy's deconvolution in the range of $R$ between
5 and 20 kpc, assuming
a ratio of thick disk stars in the solar neighborhood of $f=0.09\pm 0.01$.
With a second-order polynomial fit to the data of these scale heights and their errors
in \citet[Table 2]{Chr22}, we get
\begin{equation}
\label{flare}
h_{z,{\rm thin,G<19}}(R)=0.14-0.0037R[{\rm kpc}]+ 0.0017R[{\rm kpc}]^2 \ \ {\rm kpc}
\end{equation}\[
h_{z,{\rm thick,G<19}}(R)=1.21-0.19R[{\rm kpc}]+ 0.015R[{\rm kpc}]^2 \ \ {\rm kpc}
,\]\[
\sigma_{h_z,{\rm thin,<19}}=0.044\ {\rm kpc}; \sigma_{h_z,{\rm thick,<19}}=0.19\ {\rm kpc};
\]\[
E_{h_{z,{\rm thin,G<19}}}=|0.040-0.0080R[{\rm kpc}]
+ 0.0005R[{\rm kpc}]^2| \ \ {\rm kpc} 
,\]\[
E_{h_{z,{\rm thick,G<19}}}=|0.63-0.14R[{\rm kpc}]
+ 0.0072R[{\rm kpc}]^2| \ \ {\rm kpc} 
,\]
where $E_{h_{z,xxx}}$ stands for the error in $h_{z,xxx}$.
For the total error, we sum quadratically $E$ and the $\sigma $ of the fits of $h_z$.
These numbers were derived assuming exponential distributions with $z$, but the scale heights in the solar neighborhood are very close to those obtained of Gaia stars with sech$^2$ distributions \citep{Vie23}, and the
flare parameters affect high $z$ values where exponential and sech$^2$ distributions are equivalent.

There may be some differences between the scale height derived for the $G<19$ Gaia sample and
the stars with available radial velocity data (Gaia-RV targets; 
around 6-7\% of the stars with $G<19$ in Gaia) 
data, i.e., those with brighter magnitudes. As said previously, 
Gaia-RV targets were selected to include sources with $G<14$ covering the whole range of temperatures of the old population between the F4 and M7 spectral types of any luminosity class \citep{Kat23}.
But this difference in scale heights between both samples is small because both of them trace a 
stellar population with approximately the same age.

Thick disk stars have a very narrow range of ages, so there is no significant difference in their average age
in both surveys. Differences in the average age of the disk stars in our samples can be calculated with Besan\c con model \citep{Rob03,Amo17}.\footnote{Simulations of the Besan\c con model of stellar population synthesis 1612 are available 
at https://model.obs-besancon.fr/ } See Table \ref{Tab:besancon} for Galactic pole and anticenter directions. 
We only test it in the anticenter direction for distances less than 6 kpc because 
the Besan\c con model has a disk cutoff beyond $R=14$ kpc.
As can be observed, Gaia-RV targets have, on average, a lower age ($t$) than the complete $G<19$ survey; on average, for the ten volumes in Table \ref{Tab:besancon}, $\frac{\Delta t }{t}=-0.11$ with a dispersion of $\sigma =$0.08; $\langle t\rangle\approx 6$ Gyr.  

Scale height changes with age approximately as $\frac{\Delta h_z}{h_z}\approx 0.6\frac{\Delta t}{t}$ for $t\approx 6$ Gyr \citep{Ran92,Mac17}, so the scale height is on average 7\% (with a dispersion of 5\%) lower 
due to the change in average ages between the Gaia-RV targets and $G<19$ complete survey analyzed by \citet{Chr22},
assuming Besan\c con model. This is not an accurate calculation, and it is model dependent. Moreover, the flare
also depends on the stellar age \citep{Mac17}, which we do not take here into account. Nonetheless, we may assume in a first--order approximation for this average variation of 7\% of scale height in the thin disk with uncertainty of the dispersion for the different lines of sight; thus, our corrected scale heights for our sample of
Gaia-RV targets are
\begin{equation}
\label{flarec}
h_{z,{\rm thin}}(R)=(0.93\pm 0.05)h_{z,{\rm thin,G<19}}(R)
\end{equation},\[
h_{z,{\rm thick}}(R)=h_{z,{\rm thick,G<19}}(R)
.\]

 \begin{table}
\caption{Average age (in Gyr) of the thin disk stars,
derived from Besan\c con model for different selection of sources
according to the selected area centered at ($\ell $,$b$) with 100 deg$^2$ and heliocentric distance ($d$ in kpc); magnitude range in Gaia-G-band and spectral 
type.}
\begin{center}
\begin{tabular}{ccc}
\ & $G<19$ & $G<14$/F4-M7 \\ \hline
$b=90^\circ$, $0<d\le 2$  & 6.60 & 6.47 \\
$\ell =180^\circ $, $b=5^\circ $, $0<d\le 2$ & 4.91 & 4.32 \\
$\ell =180^\circ $, $b=5^\circ $, $2<d\le 4$ & 5.30 & 4.71  \\
$\ell =180^\circ $, $b=5^\circ $, $4<d\le 6$ & 5.35 & 3.79  \\ 
$\ell =180^\circ $, $b=15^\circ $, $0<d\le 2$ & 6.06 & 5.17 \\
$\ell =180^\circ $, $b=15^\circ $, $2<d\le 4$ & 6.95 & 6.70  \\
$\ell =180^\circ $, $b=15^\circ $, $4<d\le 6$ & 7.16 & 6.45  \\
$\ell =180^\circ $, $b=25^\circ $, $0<d\le 2$ & 6.45 & 5.68 \\
$\ell =180^\circ $, $b=25^\circ $, $2<d\le 4$ & 7.55 & 7.28  \\
$\ell =180^\circ $, $b=25^\circ $, $4<d\le 6$ & 7.76 & 6.86  \\ \hline
\end{tabular}
\end{center}
\label{Tab:besancon}
\end{table}

\subsection{Parametrization of vertical velocity rms}
\label{.rms}   

\begin{figure}
\vspace{1cm}
\includegraphics[width=8cm]{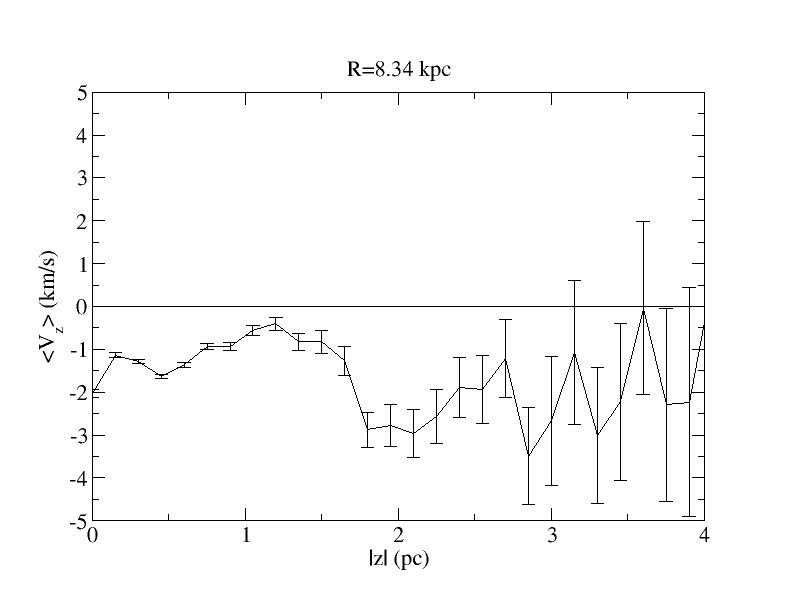}
\vspace{.2cm}
\includegraphics[width=8cm]{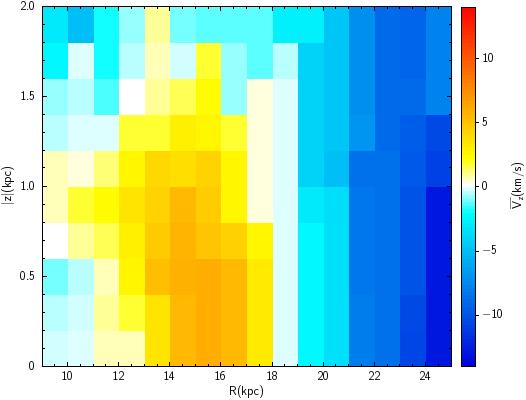}
\caption{Average $V_Z$ from Gaia DR3 extended kinematic maps within the Galactic poles (top panel) and anticenter (bottom panel) 
lines of sight.}
\label{Fig:vz}
\end{figure}

The maps of $\overline{V_z}(R,z)$ are shown in Fig. \ref{Fig:vz}.
For the rms, $\sigma _{V_z}(R,z)$, which provides a more important component in our calculations, we carry out a fit over
the data in order to obtain a smooth function, which reduces the noise in the calculation of its derivative.
We carry out fits with a fourth-order polynomial as a function of $z$ with the linear term in $z$ set to zero. 
This functional behavior is chosen as the first derivative of $\sigma _{V_z}(R,z)$ at $z=0$ should be zero (a minimum, given the symmetry for $z>0$ and $z<0$). 

\begin{figure}
\vspace{1cm}
\includegraphics[width=8cm]{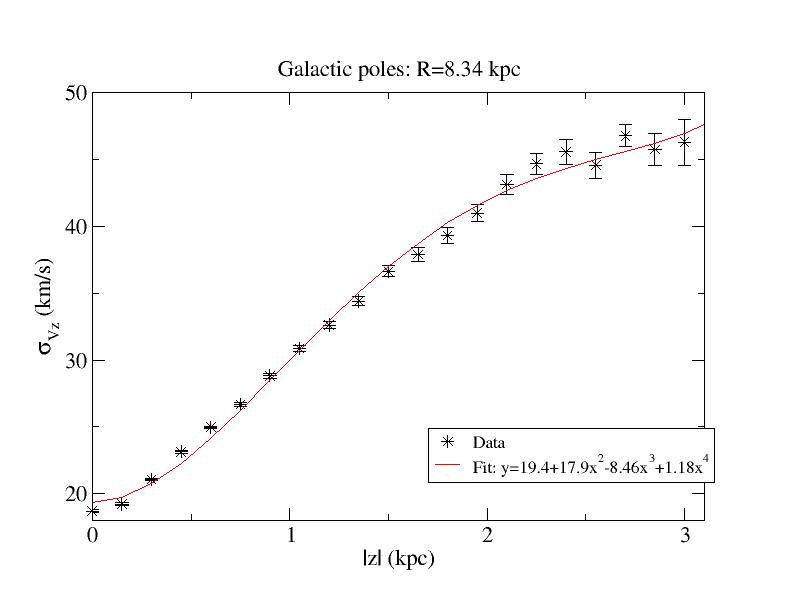}
\caption{Fit of $\sigma _{V_z}(R=R_\odot, z)$ in the Galactic pole regiond.}
\label{Fig:sigmavzpoles}
\end{figure}

\begin{figure}
\vspace{1cm}
\includegraphics[width=8cm]{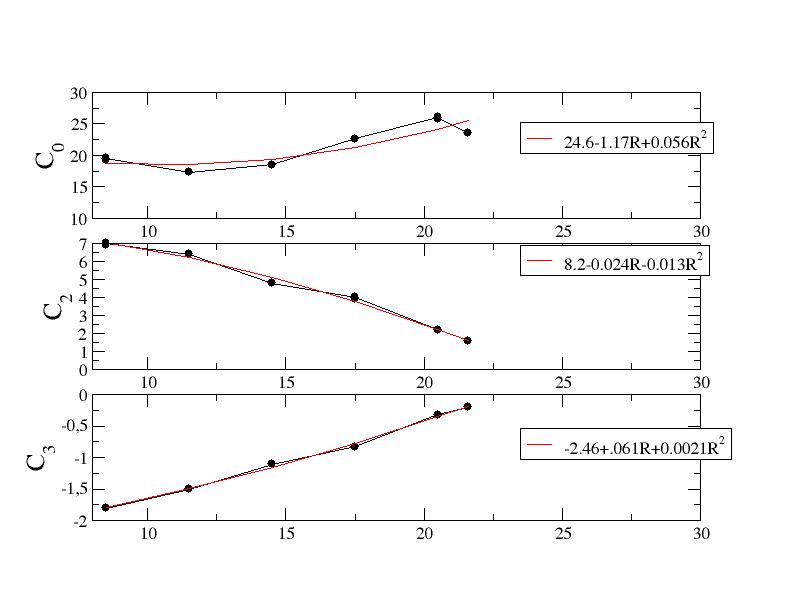}
\caption{Fit of $C_i$ in the anticenter regions. Units of $C_i$: km\,s$^{-1}$\,kpc$^{-i}$.}
\label{Fig:fitC}
\end{figure}

Our fits get the following results ($R$, $z$ in units of kpc; 
$\sigma _{V_z}$ in units of km\,s$^{-1}$):
\begin{equation}
\label{fitsigmavz}
\sigma _{V_z}(R,z)=C_0(R)+C_2(R)z^2+C_3(R)z^3+C_4(R)z^4
.\end{equation}
For the anticenter regions, we keep $C_4=0$ and we do an extra fit for each of the coefficients such that
\begin{equation}
C_i(R)=a_{0,i}+a_{1,i}R+a_{2,i}R^2
.\end{equation}
For the Galactic pole regions, we assume a unique value of $R=R_\odot$, so $a_{1,i}$ and $a_{2,i}$ are set to zero.
The values of the coefficients $a_{j,i}$ are given in Table \ref{Tab:coeffs}.
The fit of $\sigma _{V_z}(R=R_\odot, z)$ in the Galactic pole regions is shown in Fig. \ref{Fig:sigmavzpoles}.
The fits of $C_i$ to second-order polynomials in the anticenter regions
are given in Fig. \ref{Fig:fitC}.

Since the coefficients of the polynomials are correlated, rather than
calculating the errors of each coefficient in the fits,
we calculate for each value of $R$ the average deviation (rms) in the whole range of $z$ 
of $\sigma _{V_z}(R,z)$ with respect to the fit of Eq. (\ref{fitsigmavz}), 
and we fit it as a function of $R$. 
Given that the errors for the velocities contain an important ratio (unknown) of systematic errors rather than statistical ones alone, the total error is not the error in the fit but somewhat larger. There are two sources of errors: the errors in the fit (statistical) and the errors in each of the measurements of $\sigma_{V_z}^2$, which we sum quadratically as a conservative estimation. We assume the error of $\sigma _{V_z}$
is equal to the quadratic sum of the rms of the fit ($\sigma _{\rm fit}$) and the average
error in each measurement of  $\sigma _{V_z}$, where
\begin{equation}
\label{efitsigmavz}
{\rm Gal.\ poles: \sigma_{\rm fit}}[\sigma _{V_z}(R,z)]=
\end{equation}\[
0.54-0.025|z|[{\rm kpc}]+0.043z[{\rm kpc}]^2\ {\rm km/s}
\]\[
{\rm Anticenter: \sigma_{\rm fit}}[\sigma _{V_z}(R,z)]=
\]\[
2.89-0.341R[{\rm kpc}]+0.0158R[{\rm kpc}]^2\ {\rm km/s}
.\]
% ^
%The errors of the $C_i$ equal to their rms in the fits of Fig. \ref{Fig:fitC} are
%${\rm rms}[C_0]=1.42 \ {\rm km/s},
%{\rm rms}[C_2]=0.18 \ {\rm km/s/kpc}^{-2}, 
%{\rm rms}[C_3]=0.036 \ {\rm km/s/kpc}^{-3}$ in the anticenter regions.

\begin{table}
\caption{Coefficients of the fit in Eqs. (\ref{fitsigmavz}). 
Units of $a_{j,i}$: km\,s$^{-1}$\,kpc$^{-(i+j)}$.}
\begin{center}
\begin{tabular}{cccc}
$i$ & $a_{0,i}$ & $a_{1,i}$ & $a_{2,i}$ \\ \hline
Galactic poles: & & & \\
0 &  19.387 & 0 & 0  \\
2 &  17.862 & 0 & 0  \\
3 & -8.459 &  0 & 0  \\ 
4 &  1.175 & 0 & 0 \\ \hline
Anticenter: & & & \\
0 &  24.576 & -1.173 & 0.05618  \\
2 &  8.1905 & -0.023805 & -0.013001  \\
3 & -2.4622 &  0.060869 & 0.0020495  \\
 \hline
\end{tabular}
\end{center}
\label{Tab:coeffs}
\end{table}

We note that this dispersion of vertical velocities corresponds to the rms once
subtracted quadratically from the measurement errors in radial velocities and proper motions
\citep[Sect. 4]{Lop19}, so it reflects the rms of the real distribution of velocities and
not the dispersion due to errors in their measurements.

\begin{figure}
\vspace{1cm}
\includegraphics[width=8cm]{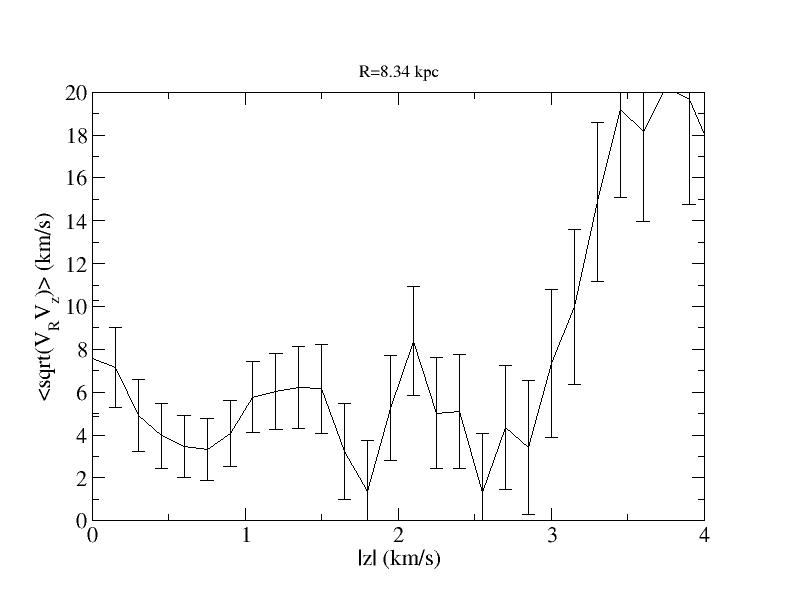}
\vspace{.2cm}
\includegraphics[width=8cm]{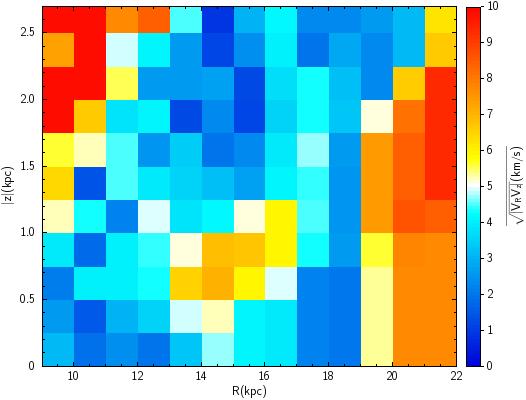}
\caption{Average $\sqrt{|V_RV_z|}$ from Gaia DR3 extended kinematic maps within the Galactic poles (top panel) and anticenter (bottom panel) lines of sight.}
\label{Fig:vrvz}
\end{figure}

We also show in Fig. \ref{Fig:vrvz} the average values of $\sqrt{|V_RV_z|}$, lower than 8 km/s for 
the regions to be explored here with significant $\Sigma $, most of them even lower than 5 km/s, 
thus giving that $|\overline{V_RV_z}|\lesssim 0.04 \sigma _{V_z}^2$ (however, in the Galactic
pole directions, for $|z|>3$ kpc, the values are larger, so we keep the exploration region within
$|z|<3$ kpc $\forall R$). Hence, and given also that $h_R\sim (3-5)\times h_z$ (see Sect. \ref{.visible}),
the term in Eq. (\ref{az}) $|\overline{V_RV_z}|\left(\frac{1}{h_R}-\frac{1}{R}\right) \lesssim 0.01\frac{\overline{V_z^2}}{h_z}$ [the dominant term  in Eq. (\ref{az})];
also the term $\frac{\partial \overline{\overline{V_RV_z}}}{\partial R}$ is much smaller than
$\frac{\partial \overline{V_z^2}}{\partial z}$ by a similar ratio.
That allows to neglect,  as we did in Sect. \ref{.dens.kin}, the terms of $V_RV_z$ in Eq. (\ref{az}) 
since it is two orders of magnitude lower than the contribution of $\sigma _{V_z}^2$ and error bars are much larger than this small $\sim 1$\% relative contribution.

\subsection{First term of Eq. (\ref{4pigro})}

From \citet{Wan23}'s rotation curves, we get $\beta (z=0)=-1.7\pm 0.1$ km s$^{-1}$kpc$^{-1}$ for $R<25$ kpc, 
and very small dependence of the slope of rotation curves with $z$ \citep[Fig. 10]{Wan23} with variations within 30\%,
so we can consider $\beta (z)\approx-1.7\pm 0.5$ km s$^{-1}$kpc$^{-1}$ $\forall z$. 
The amplitude of the rotation curve may be modeled as $V_c(R_\odot ,z)\approx 229(1+\gamma |z|)$ km s$^{-1}$,
$\gamma \sim -0.03\pm 0.01$ kpc$^{-1}$ \citep[Fig. 10]{Wan23}.

Therefore, the 
first term of the right-hand side of Eq. (\ref{4pigro}) divided by $4\pi G$ is
\begin{equation}
\Sigma _1 (R,z)=\frac{V_c(R_\odot ,z)\int_{-z}^{z} \beta (z')\mathrm{d}z'}{2\pi \,G\,R}
\end{equation}\[
\approx -(3.6\pm 1.1)[z({\rm kpc})-(0.03\pm 0.01)z({\rm kpc})^2](R_\odot /R) \ {\rm M_\odot\,pc}^{-2}
.\]

\subsection{Second term of Eq. (\ref{4pigro})}

\begin{equation}
\Sigma _2 (R,z)=-\frac{1}{2\pi \,G}\frac{\partial [\overline{V_z}^2(R,z)+\sigma _{V_z}^2(R,z)]}{\partial z}
\end{equation}

Here, we just need the information on the distribution of vertical velocities we obtained.
The errors in the vertical velocity and its rms are calculated as the average in each bin; we do not sum them 
quadratically because they are dominated by systematic errors rather than statistical ones due to uncertainties
in the distance and the application of Lucy's algorithm for the deconvolution of errors.
We calculate the derivative of $\overline{V_z}^2(R,z)$ as the average one on scales $\Delta z=$150, 300 and 450 pc, in order 
to reduce the noise; for $\sigma _{V_z}^2(R,z)$ we use the parameterization obtained in Eqs. (\ref{fitsigmavz}), (\ref{efitsigmavz}), which gives ($R$, $z$ in units of kpc; 
$\sigma _{V_z}$ in units of km\,s$^{-1}$):
\begin{equation}
\frac{\partial [\sigma _{V_z}^2(R,z)]}{\partial z}=
[4z\,C_2(R)+6z^2C_3(R)+8z^3C_4(R)]\,\sigma _{V_z}(R,z)
.\end{equation}

\subsection{Third term of Eq. (\ref{4pigro})}

\begin{equation}
\Sigma _3 (R,z)=\frac{\overline{V_z}^2(R,z)+\sigma _{V_z}^2(R,z)}{2\pi\,G\,\overline{h_z}(R,z)}
.\end{equation}
Again, we need information on the distribution of vertical velocities used for the second term, and,
moreover, we need information on the flared scale height of the combination of thin and thick disks.
The average velocity, which is a minor component, is taken directly from the data. The
$\sigma _{V_z}^2(R,z)$ and its error are taken from the parameterization obtained in Eqs. (\ref{fitsigmavz}),
(\ref{efitsigmavz}).
Eqs. (\ref{flare}), (\ref{flarec}) are used to derive the effective scale height $\overline{h_z}$ in each point through Eq. (\ref{hz}).

\subsection{Results}

\begin{figure*}
\vspace{.2cm}
\includegraphics[width=8.5cm]{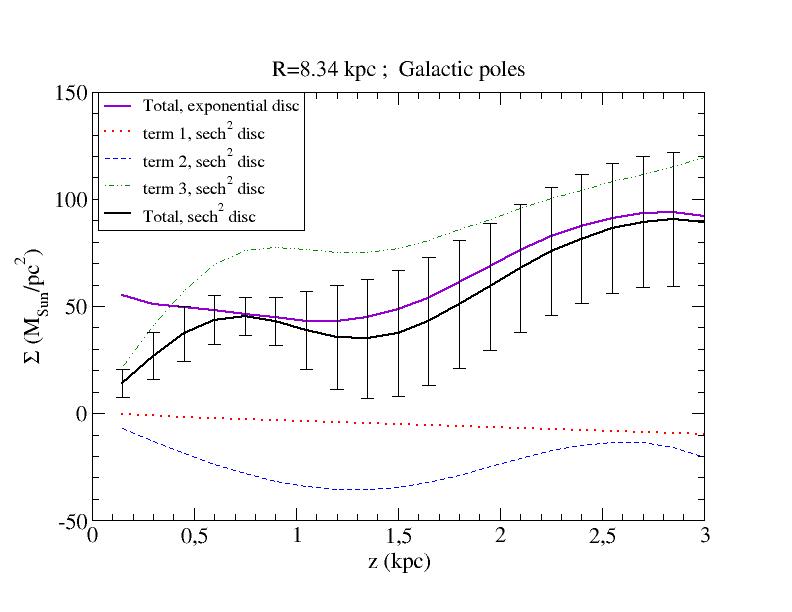}
\hspace{.2cm}
\includegraphics[width=8.5cm]{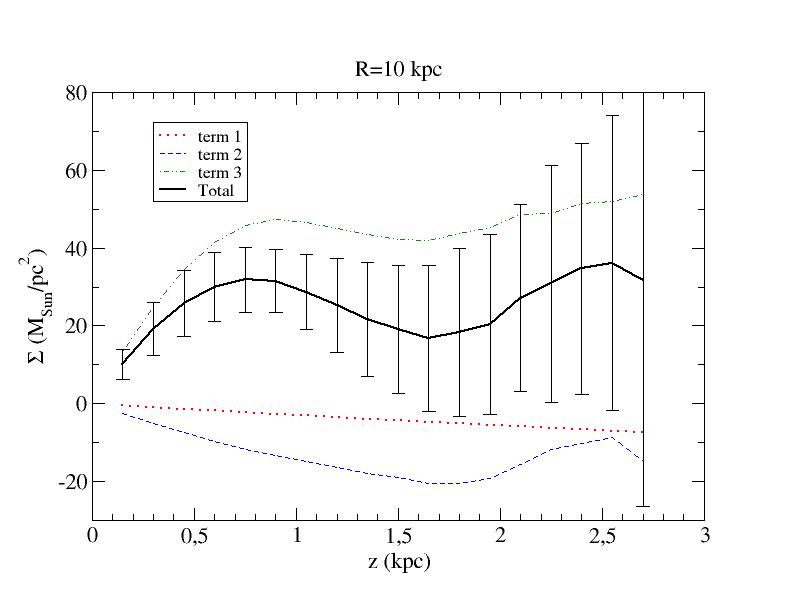}\\
\vspace{.2cm}
\includegraphics[width=8.5cm]{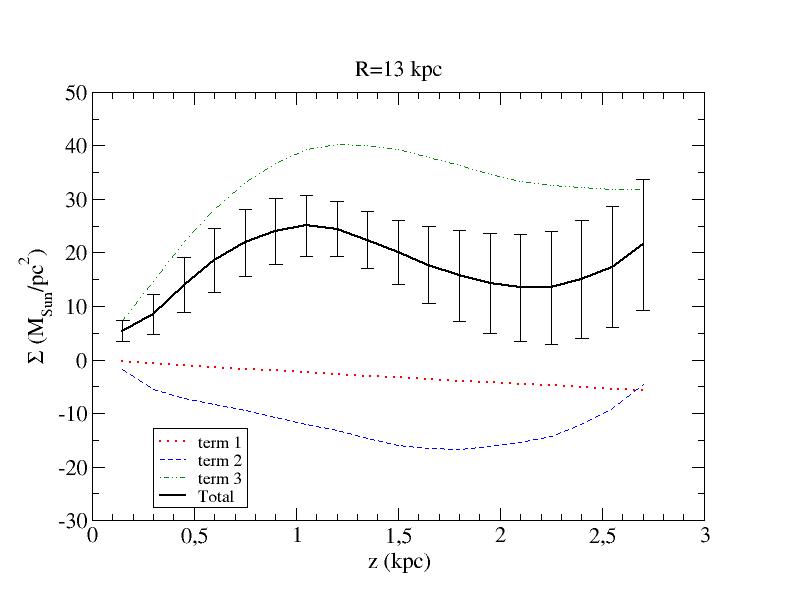}
\hspace{.2cm}
\includegraphics[width=8.5cm]{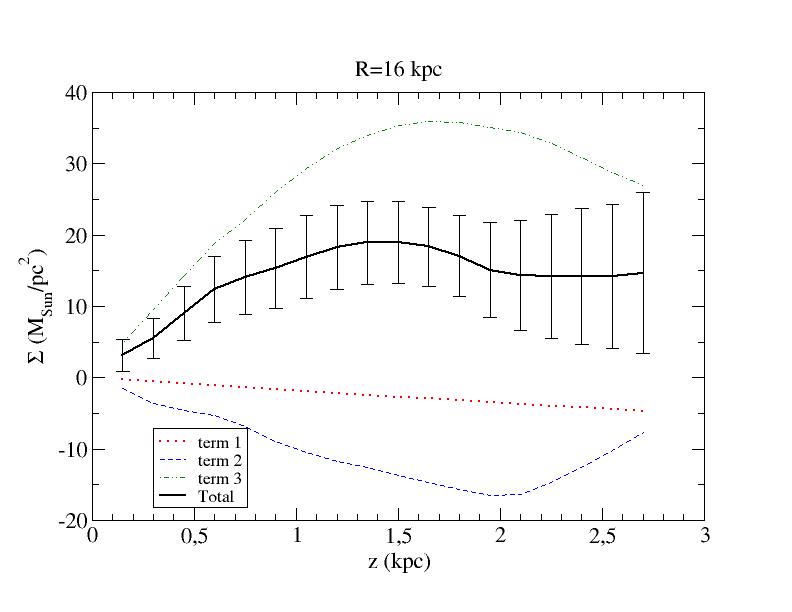}\\
\vspace{.2cm}
\includegraphics[width=8.5cm]{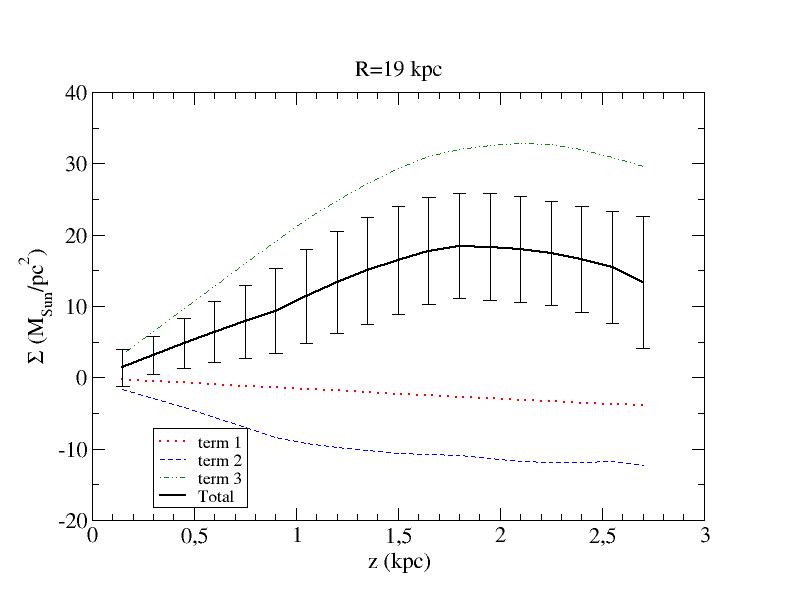}
\hspace{.2cm}
\includegraphics[width=8.5cm]{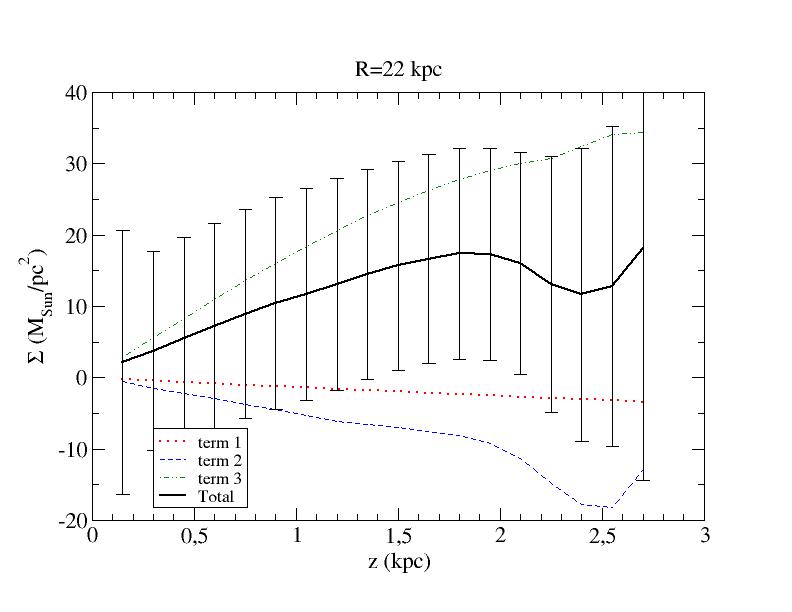}
%\hspace{.2cm}
%\includegraphics[width=8.5cm]{sigma25.jpg}
\caption{Surface density as a function of maximum $z$ for different $R$. 
Terms 1, 2 and 3 stand for the three terms ($\Sigma _1$, $\Sigma _2$ and $\Sigma _3$) that sum the total surface density.
The plot of $R=R_\odot=8.34$ kpc corresponds to the Galactic pole regions; the rest of them are from the anticenter region. Bins with $\Delta z=0.15$ kpc; $\Delta R=1$ kpc (anticenter),
$\Delta R=2$ kpc (Galactic poles). Sech$^2$ vertical density distribution for thin/thick disks 
except for $R=R_\odot $ where we also plot $\Sigma $ for exponential vertical density.}
\label{Fig:sigma}
\end{figure*}

\begin{figure}
\vspace{1cm}
\includegraphics[width=8cm]{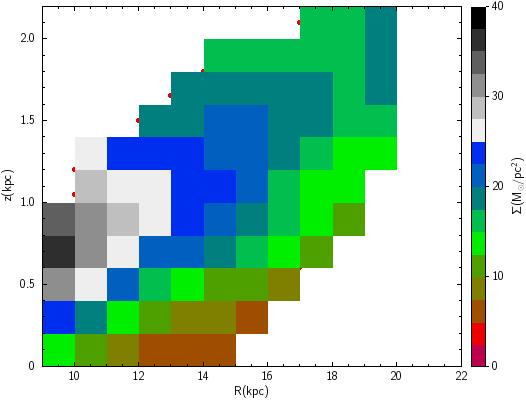}
\caption{Surface density as a function of maximum $z$ and $R$ in the anticenter. Bins with $\Delta R=1$ kpc, $\Delta z=0.15$ kpc. Only bins with
signal/noise larger than two are plotted.}
\label{Fig:sigma2D}
\end{figure}

In Fig. \ref{Fig:sigma} we show the results of the calculations of $\Sigma $ for different values of $R$.
We remind the reader that we are always using the vertical distribution as a sum of two sech$^2$ functions (Eq. \ref{rhoest});
in the case of $R=R_\odot $ we also show for comparison the case of vertical distribution as a sum of two exponentials
(Eq. \ref{rhoestexp}), in order to illustrate that $\Sigma $ near the plane is far from zero, which is inconsistent; we
should have  $\Sigma (z=0)=0$, with this being the reason to choose sech$^2$ functions in the vertical stellar distribution
throughout this paper. 
In Fig. \ref{Fig:sigma2D}, 
we plot it as a function of both $R$ and $z$ in the anticenter volume.
Due to geometric issues and the lack of stars, the relative error bars at large $|z|$ are larger than
at low $|z|$. 

At $R=R_\odot $, we get the most significant value of $\Sigma (R=R_\odot ,z=0.75\ {\rm kpc})=45\pm 9$  M$_{\odot }$ pc$^{-2}$; 
for larger $z$, the relative error bars become much larger. For instance, $\Sigma (R=R_\odot ,z=1.05\ {\rm kpc})=39\pm 18$ 
M$_{\odot }$ pc$^{-2}$. 
The origin of these large bars mainly stems from the uncertainties of $\sigma _{V_z}$ and 
$\overline{h_z}$. These results are more or less compatible with other measurements in previous works with similar methods, e.g.,
$\Sigma (R=R_\odot ,z=1\ {\rm kpc})=78.7^{+3.9}_{-4.7}$ M$_{\odot }$ pc$^{-2}$ \citep{Xia16}; 
$\Sigma (R=R_\odot ,z=1.1\ {\rm kpc})={72}_{-9}^{+6}$ M$_{\odot }$ pc$^{-2}$ \citep{Hor24}; 
$\Sigma (R=R_\odot ,z=1.1\ {\rm kpc})=55.5\pm 1.7$ M$_{\odot }$ pc$^{-2}$ \citep{Nit21};
$\Sigma (R=R_\odot ,z=1\ {\rm kpc})\approx $65 M$_{\odot }$ pc$^{-2}$ \citep{Che24}, although we
suspect that these much lower error bars are due to the fact those authors have not fully taken into account
all of the uncertainties.

At $R>R_\odot $ in the anticenter, we get results consistent with other independent
analyses \citep[Fig. 7]{Nit21}, although with much larger error bars: 
$\Sigma (R=10\ {\rm kpc},z=1.05\ {\rm kpc})=28.7\pm 9.6$ M$_{\odot }$ pc$^{-2}$;
$\Sigma (R=13\ {\rm kpc},z=1.05\ {\rm kpc})=25.0\pm 5.7$ M$_{\odot }$ pc$^{-2}$;
$\Sigma (R=16\ {\rm kpc},z=1.05\ {\rm kpc})=16.9\pm 5.8$ M$_{\odot }$ pc$^{-2}$;
$\Sigma (R=19\ {\rm kpc},z=1.05\ {\rm kpc})=11.4\pm 6.6$ M$_{\odot }$ pc$^{-2}$;
$\Sigma (R=22\ {\rm kpc},z=1.05\ {\rm kpc})=11.7\pm 14.9$ M$_{\odot }$ pc$^{-2}$.
Similar numbers were also obtained in \citet[Fig. 11/top panel]{Syl24b} (note that $a_z=2\pi G\Sigma$) 
using the same data with an
independent calculation.

\citet{Guo24} proposed a model-independent method to measure the vertical potential utilizing the intersections between the vertical phase space ($V_z$ vs. $z$), assuming that the vertical and in-plane motions are separable and the vertical energy per unit mass of the derived intersections of the
phase-space snail is conserved and connected with potential of the Galaxy. They measure the intersections of the
phase-space snail within $7.6 <R < 11.1$ kpc for Gaia DR3, and 
apply the interpolation method to deduce the potential values at several vertical heights. According to them, $\Sigma (R_\odot, z)\approx $25, 35, 50, 75 M$_\odot $/pc$^2$ for $z=$0.15, 0.30, 0.60, 1.05 kpc
respectively \citep[Fig. 12/bottom panel]{Guo24}. They are similar trends of $\Sigma $(z) to our results, 
although with slightly larger amplitude.

\subsection{Reliability of the error bars}
\label{.errors}

Although the results we obtain are roughly comparable with others obtained in the literature, we get much larger
error bars. Let us understand why the error bars are so large with an analysis for one of the values of $\Sigma $; for other
values of $R$, $z$ the analysis would be similar.

For instance, for $R=10$ kpc, $z=1.05$ kpc, we get $\Sigma =28.7\pm 9.6$ M$_{\odot }$ pc$^{-2}$. For this bin,
the three terms of Eq. (\ref{4pigro}) are: $\Sigma _1=-3.1\pm 1.0$ M$_{\odot }$ pc$^{-2}$, $\Sigma _2=-14.8\pm 1.2$ M$_{\odot }$ pc$^{-2}$, $\Sigma _3=46.6\pm 9.4$ M$_{\odot }$ pc$^{-2}$.
Clearly, the error bar of $\Sigma $ is dominated by the error in $\Sigma _3$, which is mainly produced
by the error in $\overline{h_z}(R=10\ {\rm kpc},z=1.05\ {\rm kpc})=(454\pm 51)$ pc and the error in
$\sigma _{V_z}=23.9\pm 1.7$ km/s; the contribution of $V_z$ is negligible. 
A relative error in $\sigma _{V_z}$ of 7\% makes an error of 14\% in $\sigma _{V_z}^2$, which, 
together with the error of 11\% in $\overline{h_z}$, makes the error of $\Sigma _3$ $\approx 20$\%.

Why is the relative error of $\overline{h_z}$ 11\%?
There are three sources of error in $\overline{h_z}$: (1) the error in $h_{z,{\rm thin}}=254\pm 44$ pc: this produces a relative error in $\overline{h_z}$ of 1.8\%; (2) the error in $h_{z,{\rm thick}}=810\pm 200$ pc: this produces a relative error in $\overline{h_z}$ of 10.4\%; and (3) the error in $f=0.09\pm 0.01$: this produces a relative
error in $\overline{h_z}$ of 4.1\%.

\section{Dark matter density}
\label{.dark}

\subsection{Visible matter}
\label{.visible}

The total density is the sum of the visible baryonic component ($\rho _{\rm vis}$) plus the dark matter (either baryonic or non-baryonic) component ($\rho _{\rm dark}$). 
A good knowledge of the visible matter is important to ascertain the amount of dark matter using velocity dispersions
\citep{Hes15}.
The stellar mass density  is proportional to the stellar density of the disk given
in Eq. (\ref{rhoest}), neglecting the
stellar halo contribution in the given range of $R$ and $z$, and we must add a new term for gas+dust:
\begin{equation}
\label{baryonic}
\rho _{\rm vis.}(R,z)=\rho _{M,*,\odot }\exp \left( -\frac{R-R_\odot}{h_R} \right) 
\end{equation}
\[
\times
\left[(1-f){\rm sech}^2\left(\frac{|z|}{2h_{z,{\rm thin}}(R)}\right)
+f{\rm sech}^2\left(\frac{|z|}{2h_{z,{\rm thick}}(R)}\right)\right]
\]\[
+\rho _{M,{\rm gas+dust},\odot }\exp \left( -\frac{R-R_\odot}{h_{R,{\rm gas}}}\right)
{\rm sech}^2\left(\frac{|z|}{2h_{z,{\rm gas}}(R)}\right)
\]
with the same ratio of 'thick disk'/'total disk' stars $f=0.09\pm 0.01$ \citep{Chr22}, stellar scale length
$h_R=2.2\pm 0.1$ kpc \citep{Chr22}, gas scale length $h_{R,{\rm gas}}=3.15\pm 0.60$ kpc \citep{Kal08} (error estimated by 
the dispersion on the dependence with azimuth), the stellar 
scale heights given by Eqs. (\ref{flare}) and (\ref{flarec}), and for the gas 
we assume the one derived for the HI (dominant component of gas) by \citet{Kal08} (error estimated by 
the dispersion north/south):
\begin{equation}
\label{hzgas}
h_{z,{\rm gas}}=(0.15\ {\rm kpc})\exp \left( \frac{R-R_\odot}{h_{R,fg}}\right)
,\end{equation}\[
E_{h_{z,{\rm gas}}(R)}=0.2\,h_{z,{\rm gas}},\]
$E_{h_{z,{\rm gas}}}$ stands for the error in $h_{z,{\rm gas}}$;
$h_{R,fg}=9.8$ kpc, and the normalizations in the solar neighborhood are $\rho _{M,*,\odot }= 0.033\pm 0.003$ M$_\odot $/pc$^3$, $\rho _{M,gas,\odot }=0.023\pm 0.003$ M$_\odot $/pc$^3$ in order to
get \citep{McK15}
\begin{equation}
\Sigma _{M,*,\odot }(R_\odot ,z=\infty)=
4\rho _{M,*,\odot }[(1-f)h_{z,{\rm thin}}(R_\odot )
\end{equation}\[
+f\,h_{z,{\rm thick}}(R_\odot )]=(33.4\pm 3.0) \ {\rm M}_\odot /{\rm pc}^2
,\]
\begin{equation}
\Sigma _{M,{\rm gas+dust},\odot }(R_\odot ,z=\infty)=
4\rho _{M,{\rm gas+dust},\odot }h_{z,{\rm gas}}(R_\odot )
\end{equation}\[
=(13.7\pm 1.6) \ {\rm M}_\odot /{\rm pc}^2
.\]
Errors in $\rho _{M,\odot }$ only take account the errors of $\Sigma _{M,\odot }(R_\odot ,z=\infty)$.
This makes a total density of visible matter in the solar neighborhood of 0.056$\pm 0.005$ M$_\odot $/pc$^3$
for a sech$^2$ vertical distribution, equivalent to double it, 0.112$\pm 0.010$ M$_\odot $/pc$^3$
if we assumed an exponential vertical distribution. Other
independent estimations give a similar value: for instance, $0.098^{+0.006}_{-0.014}$ M$_\odot $/pc$^3$ \citep{Gar12}
with an exponential vertical distribution.

\begin{figure}
\vspace{1cm}
\includegraphics[width=8cm]{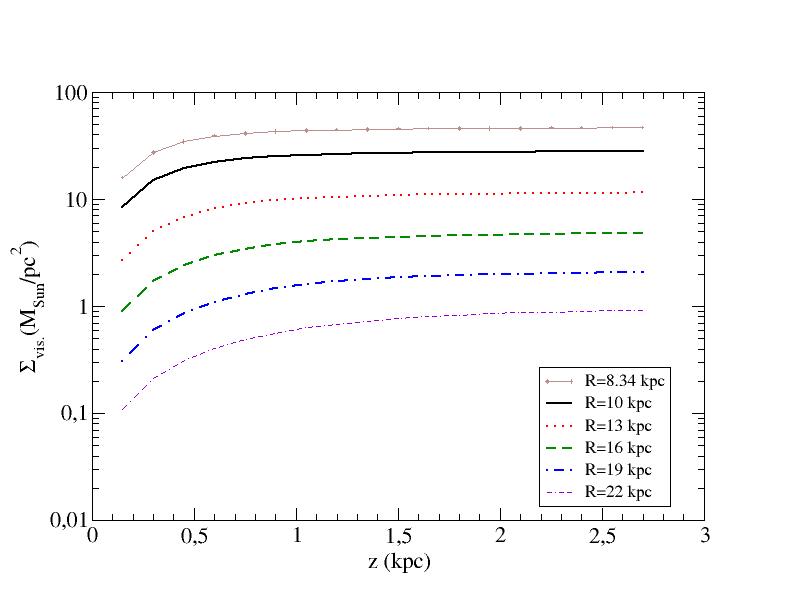}
\caption{Prediction of visible matter component of surface density as a function of maximum $z$ for different $R$.}
\label{Fig:sigmabar}
\end{figure}

%we fit $f_2$ in order to obtain a total
%$\Sigma _{\rm gas+dust}(R=R_\odot,z=\infty )=13.7$ M$_\odot $/pc$^2$ \citep{McK15}: $f_2=0.76$.
We can see the contribution of visible matter to the surface density
($\Sigma _{\rm vis.}(R,z)\equiv 
\int _{-z}^z\rho _{\rm vis.}(R,z') \mathrm{d}z'$) in Fig. \ref{Fig:sigmabar}.

\subsection{Calculation of dark matter density}

\begin{figure*}
\vspace{1cm}
\includegraphics[width=8.5cm]{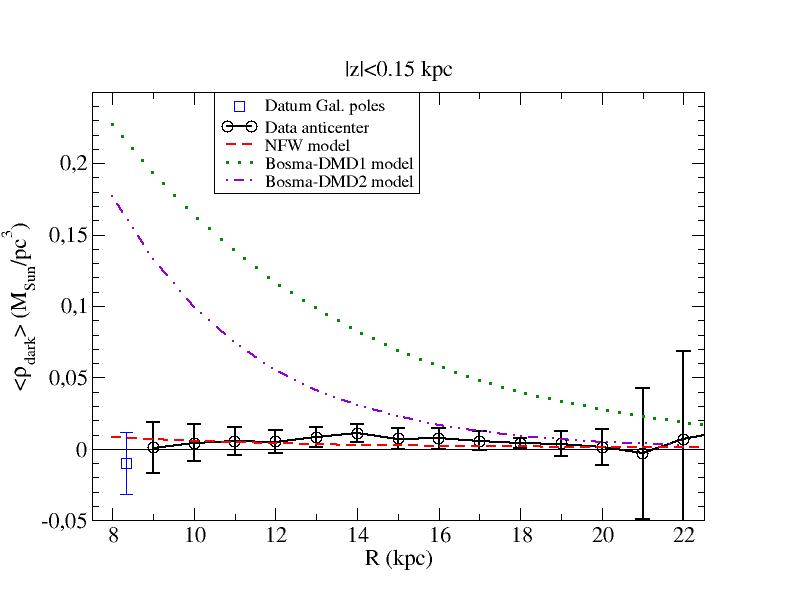}
\hspace{.2cm}
\includegraphics[width=8.5cm]{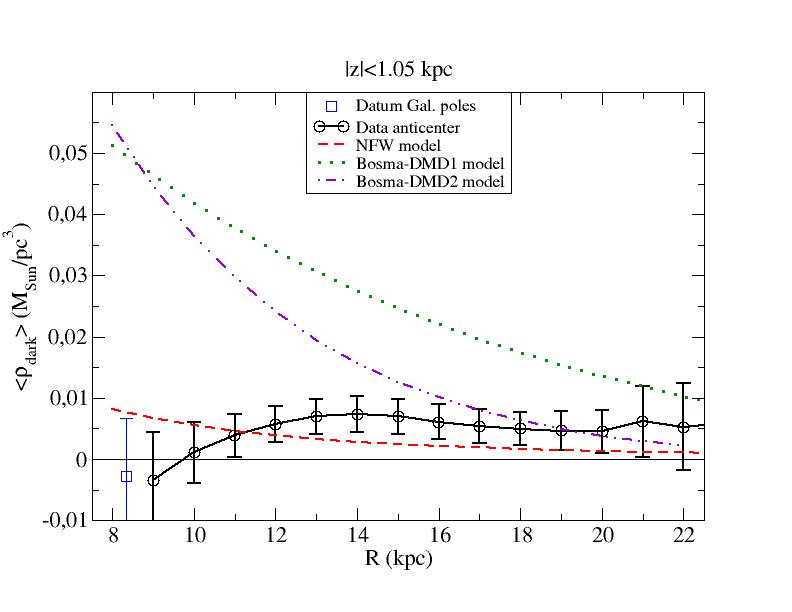}\\
\vspace{.2cm}
\includegraphics[width=8.5cm]{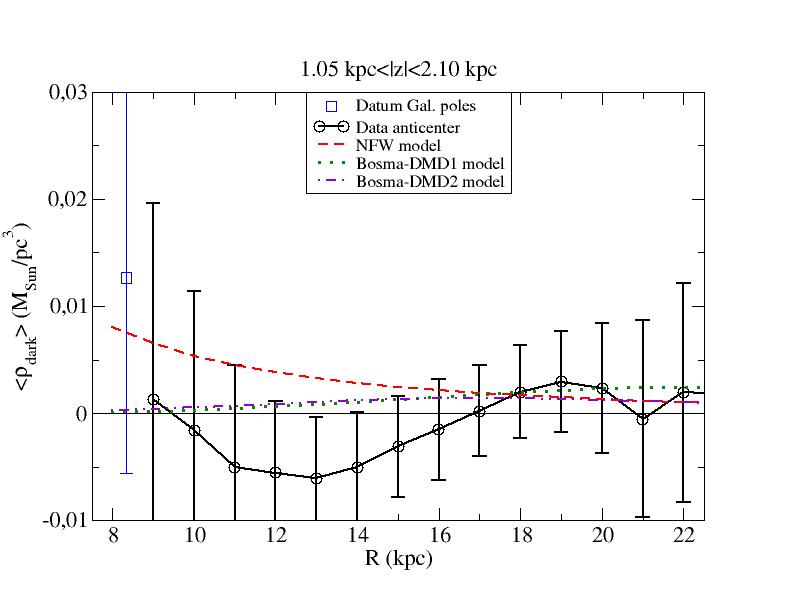}
\hspace{.2cm}
\includegraphics[width=8.5cm]{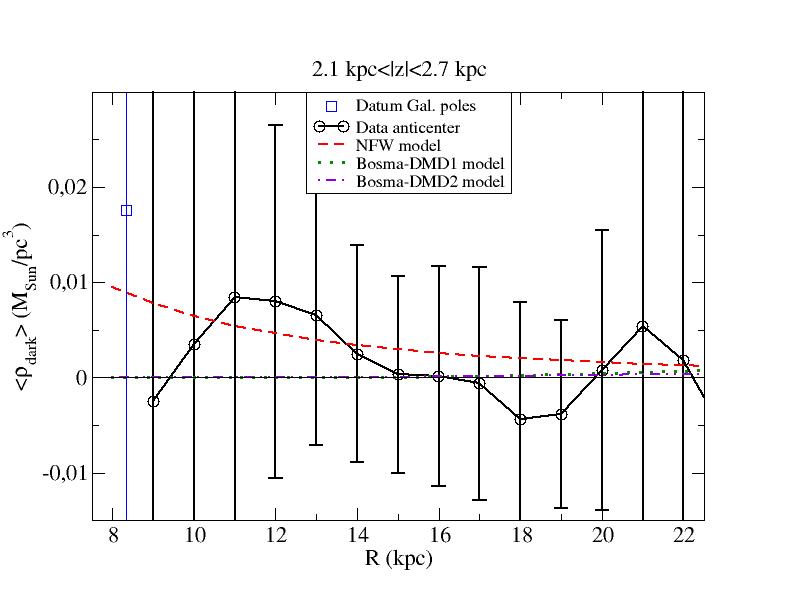}\\
\vspace{.2cm}
\includegraphics[width=8.5cm]{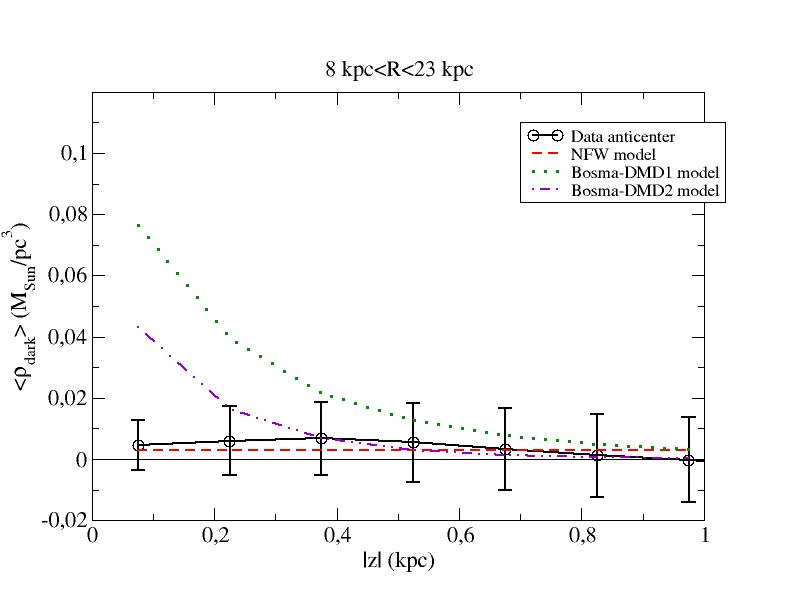}
\hspace{.2cm}
\includegraphics[width=8.5cm]{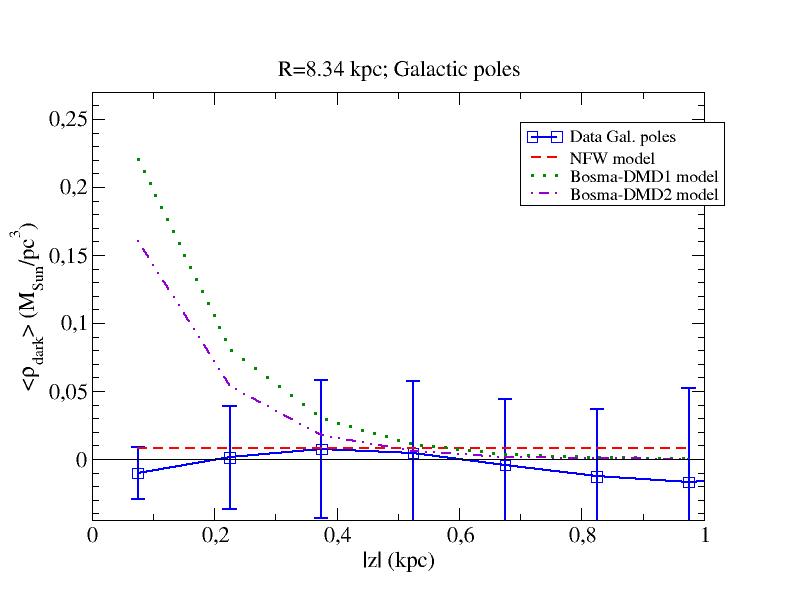}
\caption{Average dark matter density derived from Eq. (\ref{DM2}). Error bars include both the errors in the data
and the errors in the visible matter model. Top left: average within $|z|<0.15$ kpc,
as a function of $R$. Top right: average within $|z|<1.05$ kpc, 
as a function of $R$. 
Middle left: average within 1.05 kpc$<|z|<2.10$ kpc,
as a function of $R$. Middle right: average within 2.1 kpc$<|z|<2.7$ kpc, 
as a function of $R$.
Bottom left: average within 8 kpc$\le R\le 23$ kpc, as a function of $|z|$ within $|z|<1.05$ kpc in the anticenter region.
Bottom right: average density at $R=8$ kpc, as a function of $|z|$ within $|z|<1.05$ kpc in the Galactic pole regions. Comparison with theoretical models.}
\label{Fig:DM}
\end{figure*}

With this, we can calculate the dark matter density as a function of $R$ and $z$ as
\begin{equation}
\label{DM}
\rho _{\rm dark}(R,z)=\frac{1}{2}\frac{\partial [\Sigma (R,z)-\Sigma _{\rm vis.}(R,z)]}{\partial z}
,\end{equation}
or for some range with $|z|$ between $z_1$ and $z_2$,
\begin{equation}
\label{DM2}
\overline{\rho }_{\rm dark}(R,z_1<z<z_2)=
\end{equation}\[
\frac{1}{2}\frac{[\Sigma (R,z_2)-\Sigma _{\rm vis.}(R,z_2)-\Sigma (R,z_1)+\Sigma _{\rm vis.}(R,z_1)]}{z_2-z_1}
.\]
Figure \ref{Fig:DM} shows the outcome for $z_1=0$, $z_2=150$, 1050 pc; $z_1=1050$ pc, $z_2=2100$ pc; $z_1=2100$ pc, $z_2=2700$ pc.

Given the large error bars, the data are almost compatible with the total absence of dark matter.
We get an average density for $R=R_\odot $ and within $|z|<0.15$ kpc:
$\langle \rho _{\rm dark}\rangle (R=R_\odot ,|z|<0.15\ {\rm kpc})=-0.010\pm 0.022$ M$_{\odot}$ pc$^{-3}$.
Why do we get a negative value of the dark matter density? Because the model of visible matter has an
amplitude larger than the total amplitude, 
we have a total $\langle \rho _{\rm total}\rangle (R=R_\odot ,|z|<0.15\ {\rm kpc})=0.047\pm 0.021$ M$_{\odot}$ pc$^{-3}$ 
and $\langle \rho _{\rm vis.}\rangle (R=R_\odot ,|z|<0.15\ {\rm kpc})=0.056\pm 0.005$
M$_{\odot}$ pc$^{-3}$.
Nonetheless, as discussed throughout the paper, there are many uncertainties, and we are
assuming a sech$^2$ vertical distribution, which
gives a factor of two lower local density than an exponential distribution, and therefore,
a value of 0.08-0.10 M$_{\odot}$ pc$^{-3}$ for
the total density in the solar neighborhood, as measured by other groups \citep{McK15,Hor24}, 
assuming an exponential distribution, 
cannot be excluded. Indeed, we note that in our calculations
with an exponential distribution instead of sech$^2$ 
the value of $\langle \rho _{\rm total}\rangle (R=R_\odot ,|z|<0.15\ {\rm kpc})$ would rise to 0.184 M$_{\odot}$ pc$^{-3}$, so possibly some intermediate
distribution between exponential and sech$^2$ would get better values.

Again, these large error bars reflect the large uncertainties in the input parameters of the velocity dispersion and
scale height, and also the uncertainties in the models of visible matter.
The bins of the Galactic anticenter are less noisy than the ones at the Galactic poles, and for $R$ between 13 and 16 kpc we
get the most significant results. 
The hypothesis of zero dark matter [$\rho _{\rm dark}(R,z)=0\ \forall R,z$] is not far from the data.
At present, exploring the different values of $R$, $z$, the most significant departure from $\rho _{\rm dark}(R,z)=0$ is only
at 2.6$\sigma $: $\langle \rho _{\rm dark}\rangle (R=14\ {\rm kpc},|z|<1.05\ {\rm kpc}) =(7.4\pm 2.9)\times 10^{-3}$ M$_\odot $/pc$^3$.

\subsection{Comparison with dark matter models}
\label{.compDM}

From the rotation curves at $z=0$ derived from the same data of Gaia DR3 with Lucy deconvolution of 
errors, there are two possible solutions for the dark matter distribution \citep{Syl23}:

\begin{enumerate}
\item A spherical distribution with a Navarro-Frenk-White (NFW) profile:
\begin{equation}
\rho _{\rm dark,sph}(R,z)=\frac{\rho _{ds0}}{\left(1+\frac{\sqrt{R^2+z^2}}{r_s}\right)^2\frac{\sqrt{R^2+z^2}}{r_s}}
,\end{equation}
with best-fit parameters \citep[Table 2/DR3+]{Syl23}: $\rho _{ds0}=0.014$ M$_\odot $/pc$^3$, $r_s=12.6$ kpc.
This value of the density of dark matter in the solar neighborhood, $\rho _{ds0}$, is similar
to the values obtained by different teams: for instance, $0.022^{+0.015}_{-0.013}$ M$_\odot $/pc$^3$ \citep{Gar12}:
from Milky Way rotation curve implies  $\rho _{ds0}=$0.4 GeV/c$^2$ cm$^{-3}$=0.011 M$_\odot $pc$^{-3}$ 
\citep{Cat12,Pat15,deS21,Ou24}; $0.018\pm 0.0054$ M$_{\odot}$ pc$^{-3}$ \citep{Xia16}.
These parameters were derived assuming a more simple expression of visible matter, which we use here, and $\rho _{ds0}$, $r_s$ 
may slightly change for a fit of the rotation curve using Eqs. (\ref{baryonic}) instead, but these variations are small
and do not affect our analysis, which, as said below, tries to determine whether this NFW profile is compatible with
the data with some error bars much larger than these small changes due to different fits of the rotation curve.

\item A disky distribution of dark matter tracing the gas \citep{Bos81,Syl23}:

[DMD1:]
\begin{equation}
\Sigma _{\rm dark,disc}(R,\infty)=\Sigma _{dd0}\exp \left( -\frac{R}{R_d}\right)  
,\end{equation}
with best-fit parameters \citep[Table 3/DR3+]{Syl23}: $\Sigma _{dd0}$=230 M$_\odot $/pc$^2$,
$R_d=10.2$ kpc. We can assume a vertical exponential distribution; thus,
\begin{equation}
\label{discdark}
\rho _{\rm dark,disc}(R,z)=\frac{\Sigma _{dd0}}{2h_{z,{\rm dark}}(R)}\exp \left( -\frac{R}{R_d} \right)
\end{equation}\[\times
\exp\left(-\frac{|z|}{h_{z,{\rm dark}}(R)}\right)
.\]
We assume that $h_{z,{\rm dark}}(R)=h_{z,{\rm gas}}(R)$ given by Eq. (\ref{hzgas}).
Hereafter, we refer as DMD1 this model with these parameters.

%\[
%h_{z,{\rm gas}}(R)=0.106-0.0028R({\rm kpc})+ 0.0013R({\rm kpc})^2 \ \ {\rm kpc} 
%\]
%with $\rho _{dd0}=1.47$ M$_\odot $/pc$^3$ M$_\odot $/pc$^3$, $R_d=10.2$ kpc.

[DMD2:] The last parameters were obtained from rotation curves with simpler than the assumptions herein 
about the visible matter.
If we used the same model of visible matter that we used here, Eq. (\ref{baryonic}), in the fit of the rotation curve,
and also assuming sech$^2$ vertical distribution of dark matter, we would get
\begin{equation}
\label{discdark2}
\rho _{\rm dark,disc}(R,z)=\frac{\Sigma _{dd0}}{4h_{z,{\rm dark}}(R)}\exp \left( -\frac{R}{R_d} \right)
\end{equation}\[\times
{\rm sech}^2\left(\frac{|z|}{2h_{z,{\rm dark}}(R)}\right)
,\]
with $\Sigma _{dd0}$=570 M$_\odot $/pc$^2$, $R_d=5.0$ kpc (Sylos Labini, priv. comm. 2024; 
\citet{Syl24b} obtains the same values for an exponential vertical distribution too). Again, 
we assume that $h_{z,{\rm dark}}(R)=h_{z,{\rm gas}}(R)$ given by Eq. (\ref{hzgas}).
Hereafter we refer as DMD2 this model with these parameters.
\end{enumerate}

\begin{figure*}
\vspace{1cm}
\includegraphics[width=8.5cm]{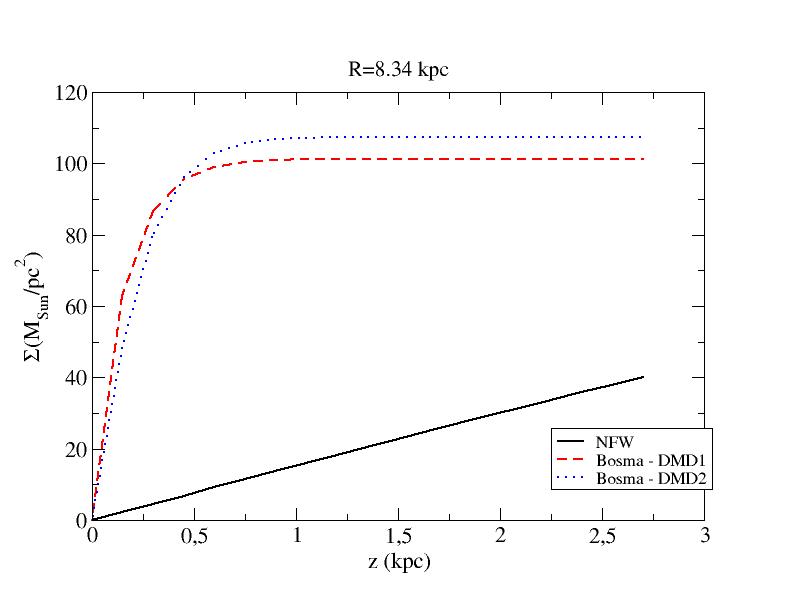}
\hspace{.2cm}
\includegraphics[width=8.5cm]{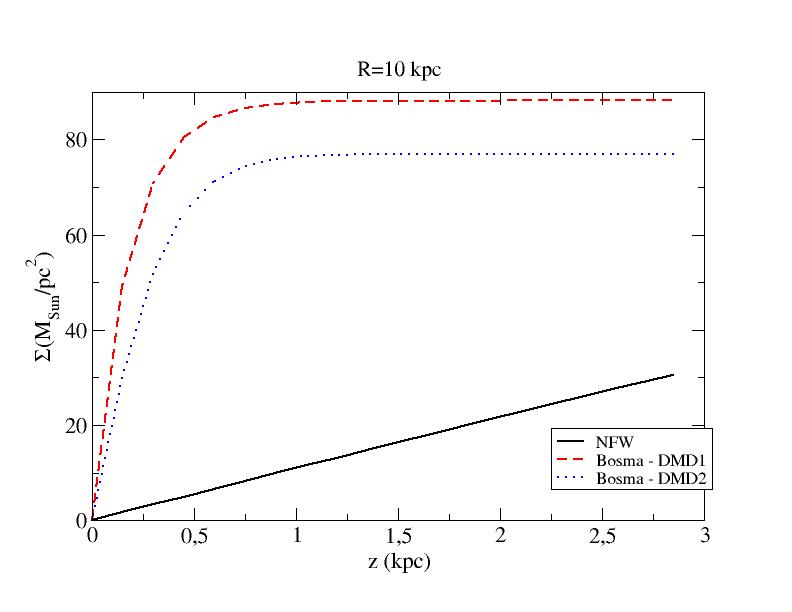}\\
\vspace{.2cm}
\includegraphics[width=8.5cm]{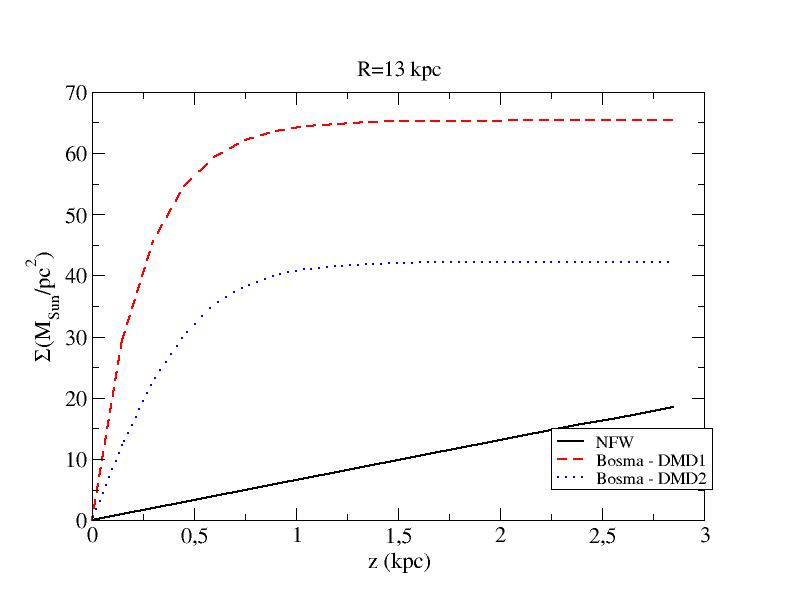}
\hspace{.2cm}
\includegraphics[width=8.5cm]{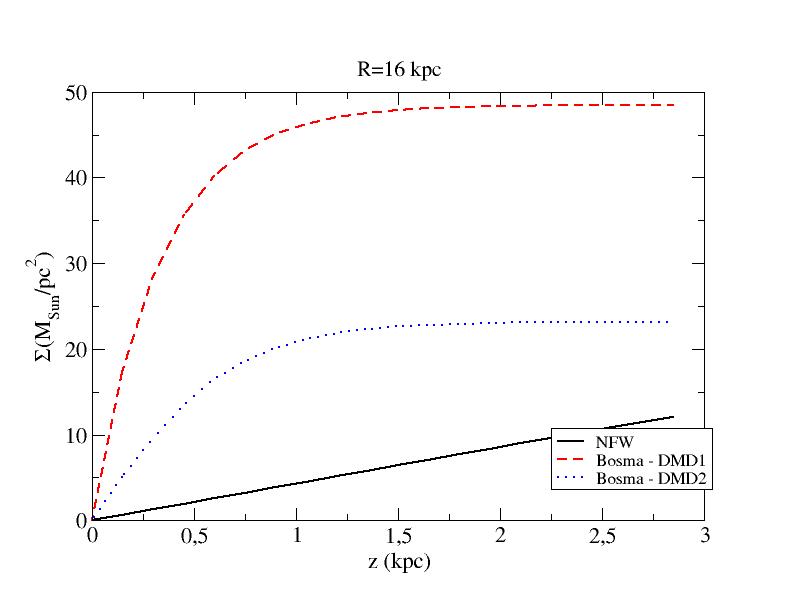}\\
\vspace{.2cm}
\includegraphics[width=8.5cm]{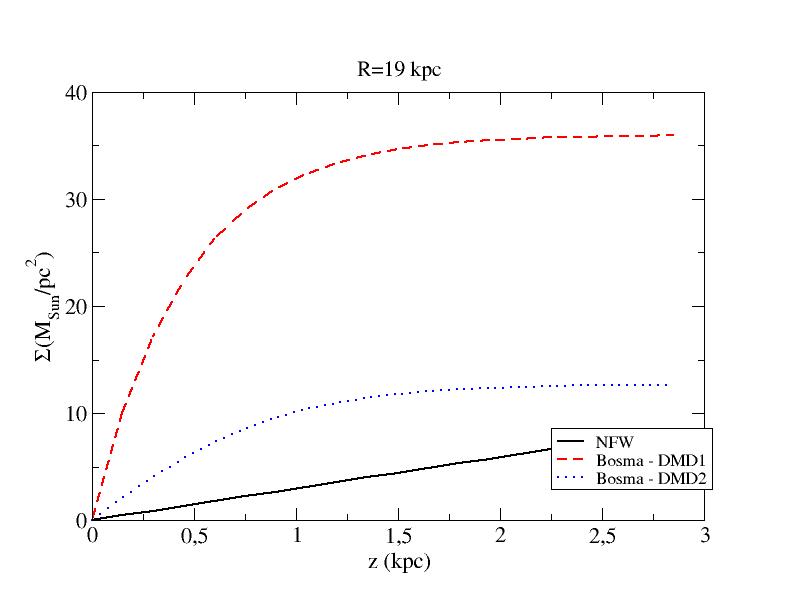}
\hspace{.2cm}
\includegraphics[width=8.5cm]{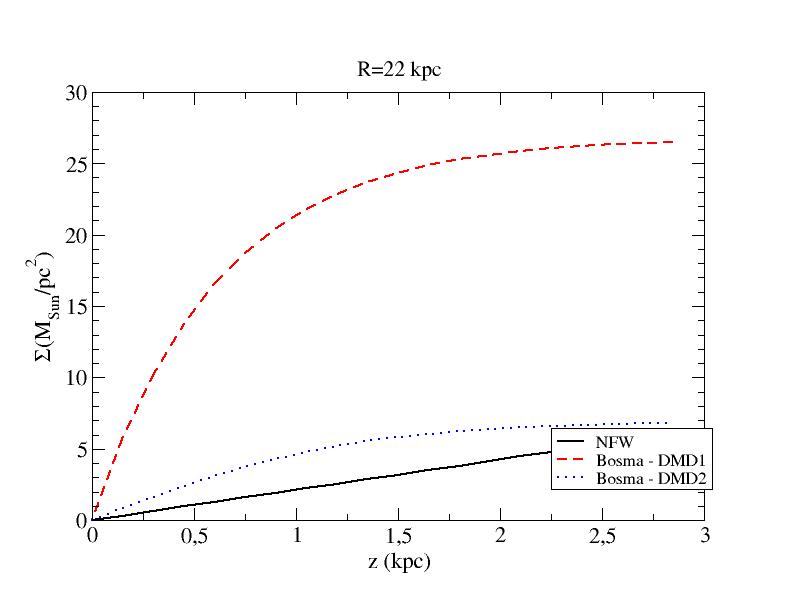}
%\hspace{.2cm}
%\includegraphics[width=8.5cm]{sigma25p.jpg}
\caption{Predicted dark matter surface density as a function of maximum $z$ for different $R$ from the models of
spherical distribution with an NFW profile and disky distribution of dark matter assuming a scale height like that of the gas.}
\label{Fig:DMpred}
\end{figure*}

Figs. \ref{Fig:DM} and \ref{Fig:DMpred} show the expected average $\rho $ or $\Sigma $ of the NFW model and the
Bosma models (DMD1 or DMD2).
%From the comparison of $\Sigma $ between data and the two models, the fit with NFW with
%$\chi ^2=346$ for $N=72$ degrees of freedom, whereas for the disky dark matter model assuming a scaleheight like the gas, 
%the fits improves very slightly, $\chi ^2=229$ for $N=72$. However,
%beware that our data are not totally independent
%and their error bars is a sum of statistical and systematic errors, which might not be well represented by a Gaussian
%distribution, and a $\chi ^2$ test may give an estimation of the best fit but not about its confidence level.
%Note also that there are no free parameters in these models; the parameters were already
%fitted with rotation curves in previous analysis or included in the fit of the gas scaleheight. 

Visually examining Fig. \ref{Fig:DM}, the NFW model is compatible with the data.

\begin{figure}
\vspace{1cm}
\includegraphics[width=8cm]{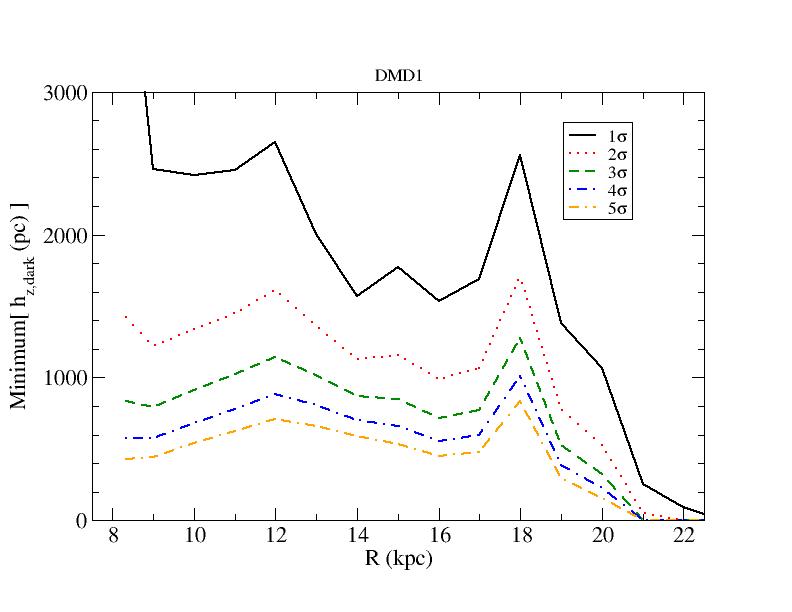}\\
\vspace{.2cm}
\includegraphics[width=8cm]{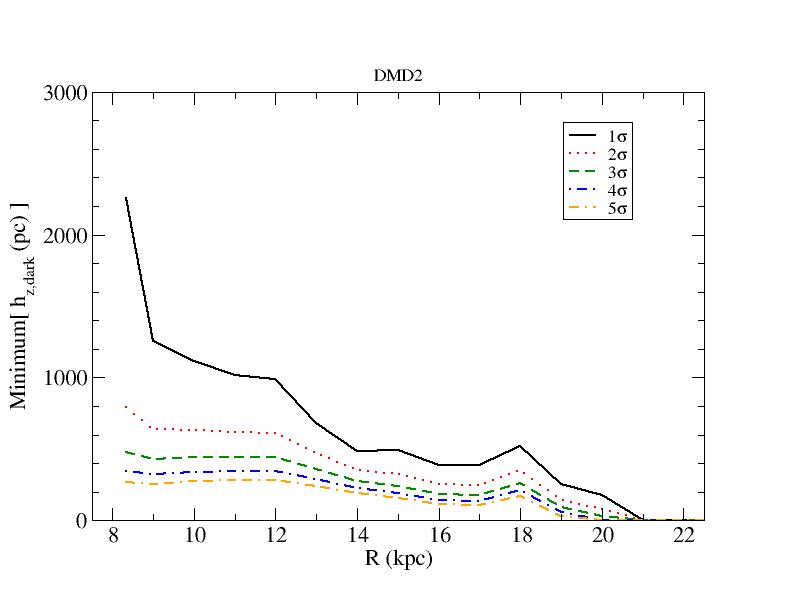}
\caption{Minimum scale height of a disky distribution of dark matter compatible with the data within 1$\sigma$-5$\sigma $. Top panel: DMD1 model;
bottom panel: DMD2 model}
\label{Fig:Bosmahz}
\end{figure}

The disky dark matter model is not compatible with the data
when we assume the dark matter is proportional to the gas, as originally proposed by Bosma.
A disky dark matter would need a much higher scale height.
Taking the data with the measurements at $R_\odot $, $|z|<0.15\ {\rm kpc}$, and comparing them
with Eq. (\ref{discdark}) or (\ref{discdark2}) with $h_{z,{\rm dark}}(R)$ as free parameter for each $R$, we get 
the minimum scale heights compatible with the data within 1$\sigma$-5$\sigma$. They are plotted in Fig. \ref{Fig:Bosmahz}.
We can see that, even assuming a 2$\sigma $ deviation in $\rho _{\rm dark}(R,|z|<0.15\ {\rm kpc})$, we need
$h_{z,{\rm dark}}>600$ pc for 8 kpc$<R<12$ kpc in the best of the cases (DMD2). 
Such a large thickness is very different from the scale height of the gas equal to 
150 pc at $R_\odot $ as expected from the Bosma hypothesis.

%If we concentrate in the data for $|z|<750$ pc, it is clear from the comparison of Figs. \ref{Fig:DM} and \ref{Fig:DMpred2}
%that NFW is unable to explain the high average density in this region (because most of the halo dark matter mass is at $|z|>750$ pc), whereas for disky dark matter profile most of the mass is concentrated at $|z|<750$ pc. The $\chi ^2$ for the comparison of $\rho $
%between data and models of Figs. \ref{Fig:DM} and \ref{Fig:DMpred2} $\forall z$
%gives $\chi ^2=142$ for $N=54$ for NFW, and $\chi ^2=225$ for the same $N=54$ for the disky dark matter; 
%again not a perfect fit, but much better than the spherical dark matter halo. 

%A distribution of $\rho _{\rm dark }$ compatible with the measured $\Sigma $ derived from vertical velocities and with the rotation curve
%still may need further refinement in the parameters, but, taking into account that the calculated error bars are approximations and
%that there are covariance terms that we have not taken into account in the calculation of $\chi ^2$, the Bosma model of dark matter
% may be not very far from reality, whereas the spherical distribution of NFW is very far from fitting $\Sigma $ measurements.

\section{Calculations with MOND}
\label{.MOND}

The calculation with MOND, which posits a Galaxy only with visible matter 
(without dark matter component) and with modifications to the gravity, differs from the previous analysis
in the variation of the Poisson equation, which now becomes \citep{Bek84,Lop21,Zhu23}
\begin{equation}
4\pi G\Sigma(R,z)=\frac{1}{R}\int_{-z}^{z} \frac{\partial [R\,a_R(R,z)\mu (R,z)]}{\partial R} \mathrm{d}z'
\end{equation}\[
+[a_z(R,z)\mu (R,z)-a_z(R,-z)\mu (R,-z)]
,\]
\[
\mu(R,z)=\left[1+\frac{a_0^2}{a_R(R,z)^2+a_z(R,z)^2}\right]^{-1/2}
,\]
where $a_0=1.2\times 10^{-10}$ m\,s$^{-2}$; and setting $\rho _{\rm dark}(R,z)=0\ \forall R\ \forall z$. We develop
this expression as we did in section \ref{.dens.kin}, considering here the approximation of $\beta =0$ (flat rotation curve).
We have seen in previous sections that the first term of $\Sigma (R,z)$ of Eq. (\ref{4pigro}) gives a negligible contribution given the low value of $\beta$, so we can neglect it:
For instance, this term gives a contribution for $\Sigma(z=1\ {\rm kpc})=3.5\frac{R_\odot}{R}$ M$_\odot $/pc$^2$, which is much smaller than the error bars, 
ranging between 16 M$_\odot $/pc$^2$  for $R=R_\odot $ to a minimum of 5.3 M$_\odot $/pc$^2$ for $R=16$ kpc, 
so it is always lower than 1/3 of the total error bar. Neglecting the term with $\beta $ should thus produce a systematic error lower than 1/3 of the total errors of other terms, which can be considered negligible.
With this one and previous approximations we get
\begin{equation}
\label{4pigroMOND}
4\pi G\Sigma(R,z)\approx \frac{V_c(R_\odot ,z)^2}{R}\int_{-z}^{z} \frac{\partial\mu (R,z')}{\partial R}\mathrm{d}z'
\end{equation}\[
-2\mu (R,z)\frac{\partial \overline{V_z^2}(R,z)}{\partial z}
+2\mu (R,z)\frac{\overline{V_z^2}(R,z)}{\overline{h_z}(R,z)}
,\]\[
\mu(R,z)=\left[1+\frac{a_0^2}{\frac{V_c(R_\odot ,z)^4}{R^2}+\left[-\frac{\partial \overline{V_z^2}(R,z)}{\partial z}+\frac{\overline{V_z^2}(R,z)}{\overline{h_z}(R,z)}\right]^2}\right]^{-1/2}
.\]

\begin{figure}[htb]
\vspace{1cm}
\includegraphics[width=8cm]{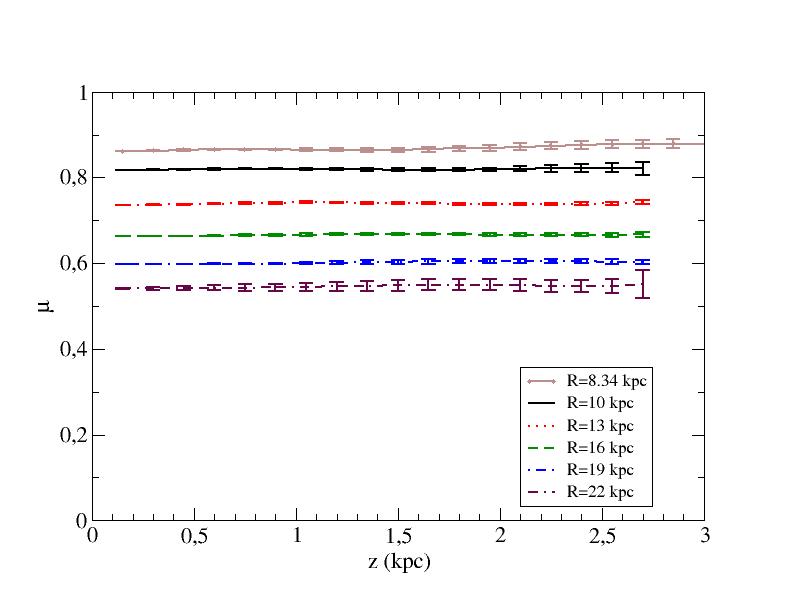}
\caption{Factor $\mu $ in MOND equations derived from the kinematical data.}
\label{Fig:muMOND}
\end{figure}

\begin{figure*}
\vspace{.2cm}
\includegraphics[width=8.5cm]{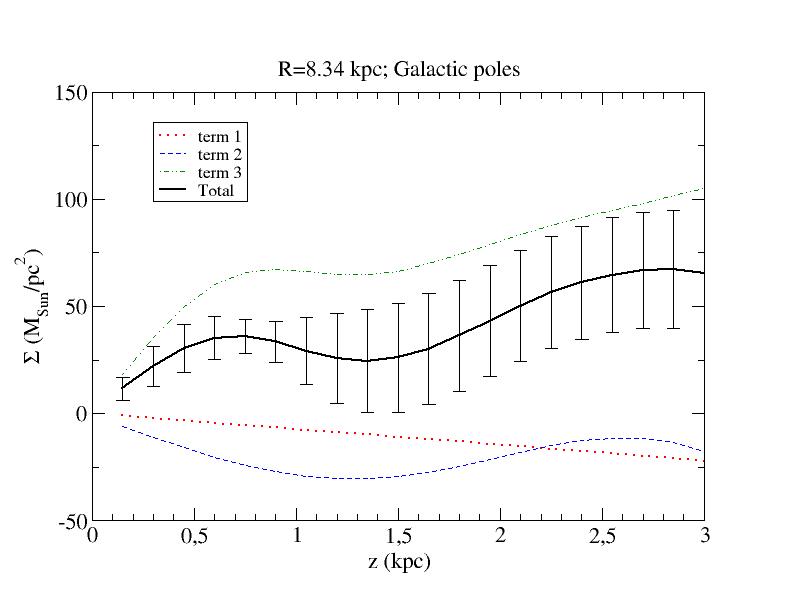}
\hspace{.2cm}
\includegraphics[width=8.5cm]{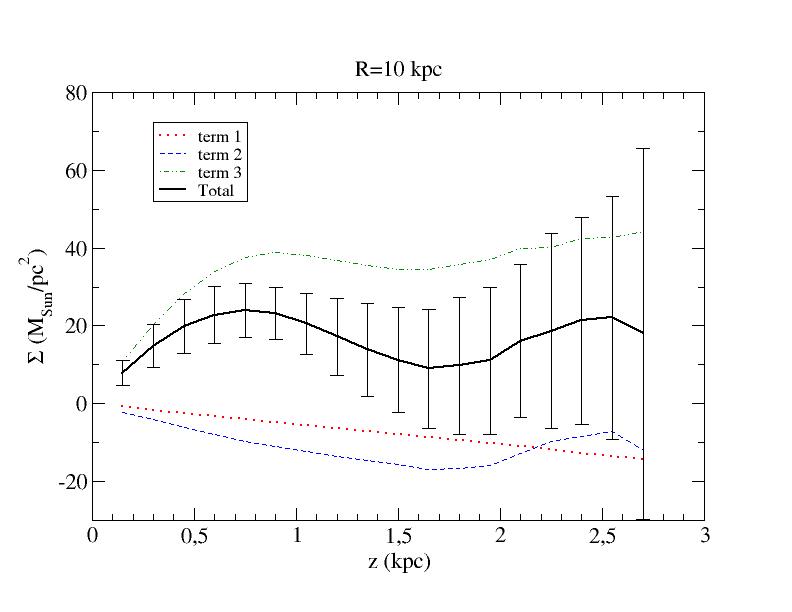}\\
\vspace{.2cm}
\includegraphics[width=8.5cm]{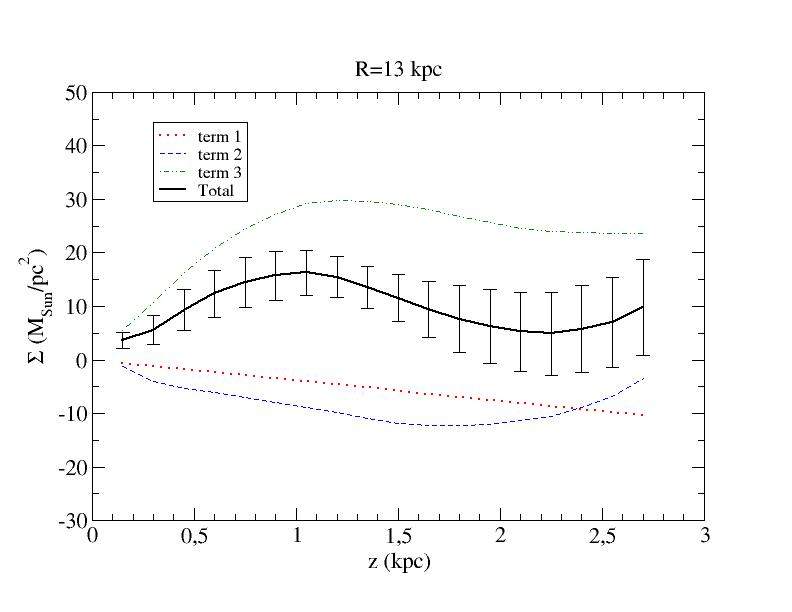}
\hspace{.2cm}
\includegraphics[width=8.5cm]{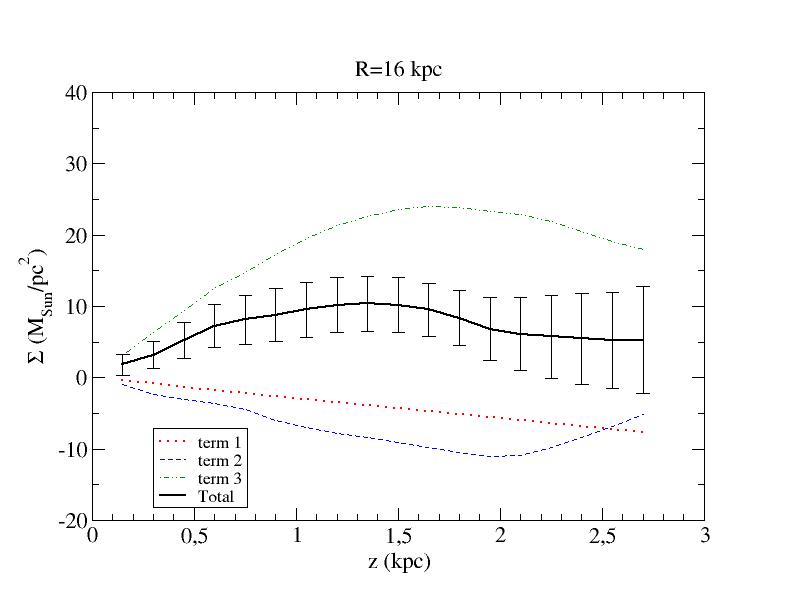}\\
\vspace{.2cm}
\includegraphics[width=8.5cm]{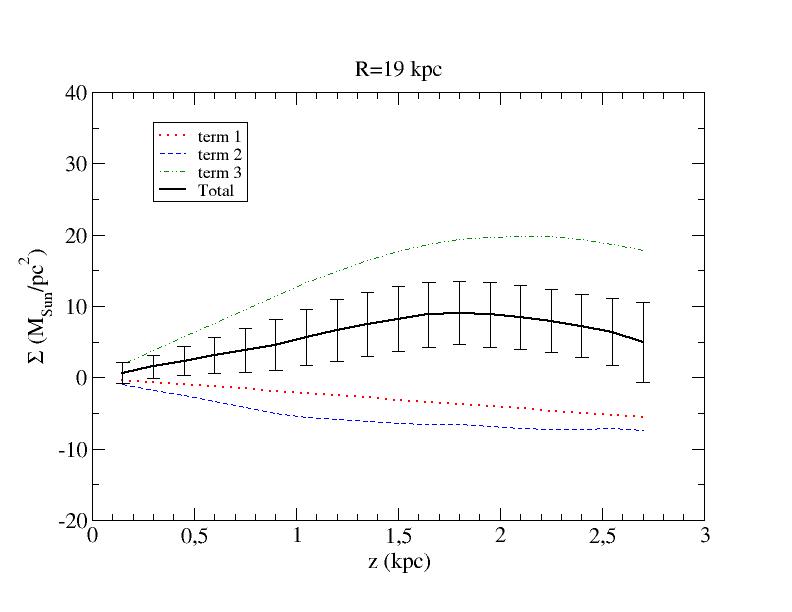}
\hspace{.2cm}
\includegraphics[width=8.5cm]{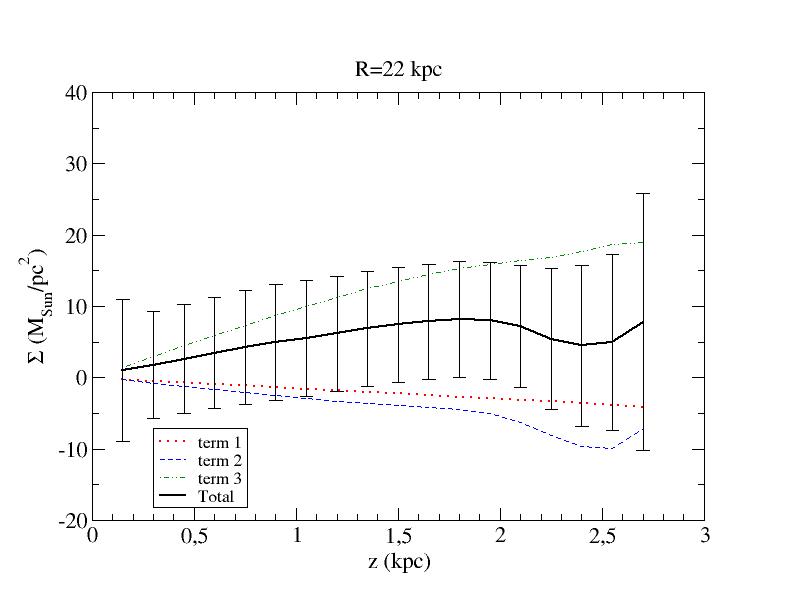}
%\hspace{.2cm}
%\includegraphics[width=8.5cm]{sigma25m.jpg}
\caption{Surface density as a function of maximum $z$ for different $R$ with MOND. 
Terms 1, 2, 3 stand for the three terms that sum the total surface density in Eq. (\ref{4pigroMOND}).
The plot of $R=8.34$ kpc corresponds to the Galactic pole regions; the rest of them are from the anticenter region. Bins with $\Delta z=0.15$ kpc; $\Delta R=1$ kpc (anticenter),
$\Delta R=2$ kpc (Galactic poles).}
\label{Fig:sigmam}
\end{figure*}

The factor $\mu$ depends only on the acceleration, and it can be derived directly from kinematics data. In Fig. \ref{Fig:muMOND},
we show its value. In Fig. \ref{Fig:sigmam}, we give the calculation of $\Sigma (R,z)$ for MOND. The first term is again
small.
We are within a moderate MONDian regime of $\mu \gtrsim 0.6$ for $R<20$ kpc, where 
$\Sigma _{MOND}(R,z)\approx \Sigma _{Newton}(R,z)\mu (R,z)$, which gives a decrease in 
the density lower than 67\% with respect to the Newtonian one.

\begin{figure*}
\vspace{1cm}
\includegraphics[width=8.5cm]{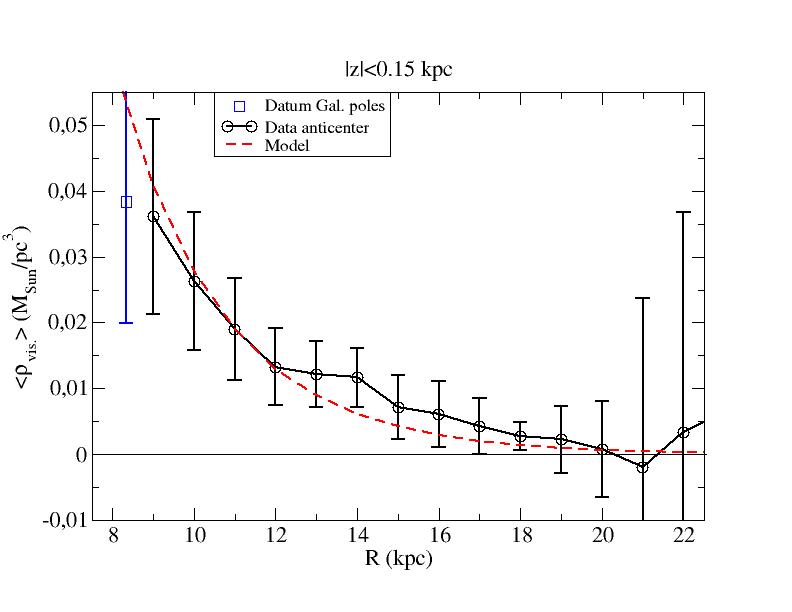}
\hspace{.2cm}
\includegraphics[width=8.5cm]{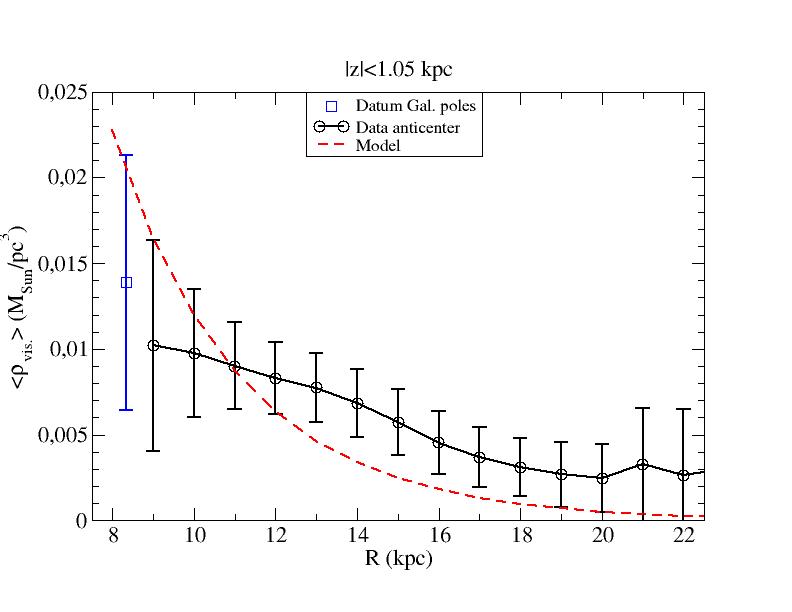}\\
\vspace{.2cm}
\includegraphics[width=8.5cm]{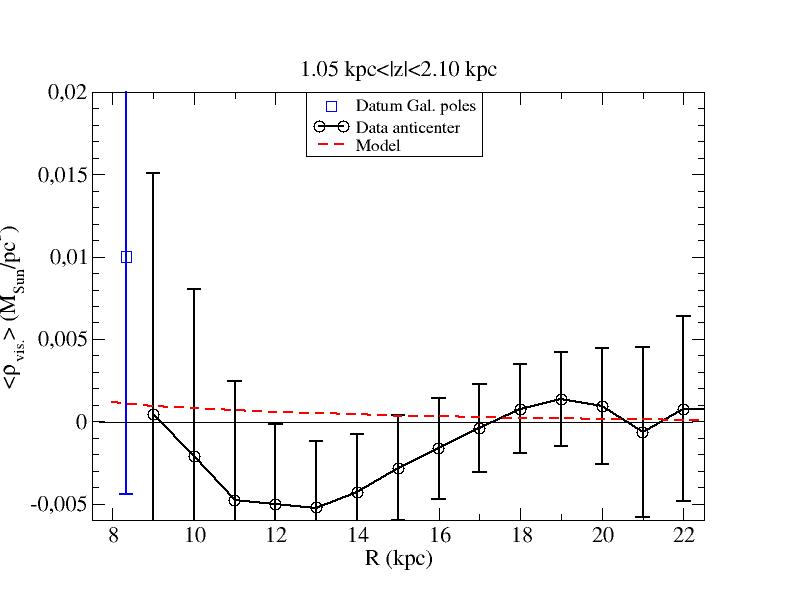}
\hspace{.2cm}
\includegraphics[width=8.5cm]{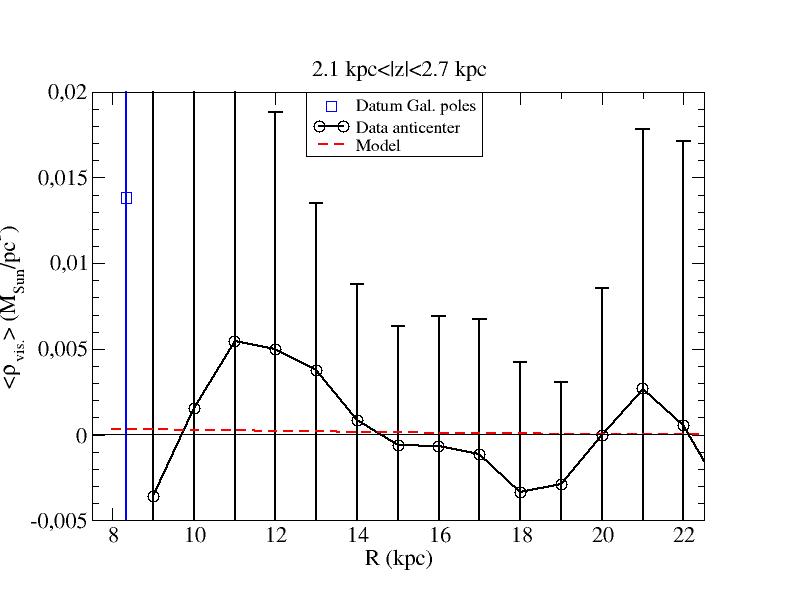}\\
\vspace{.2cm}
\includegraphics[width=8.5cm]{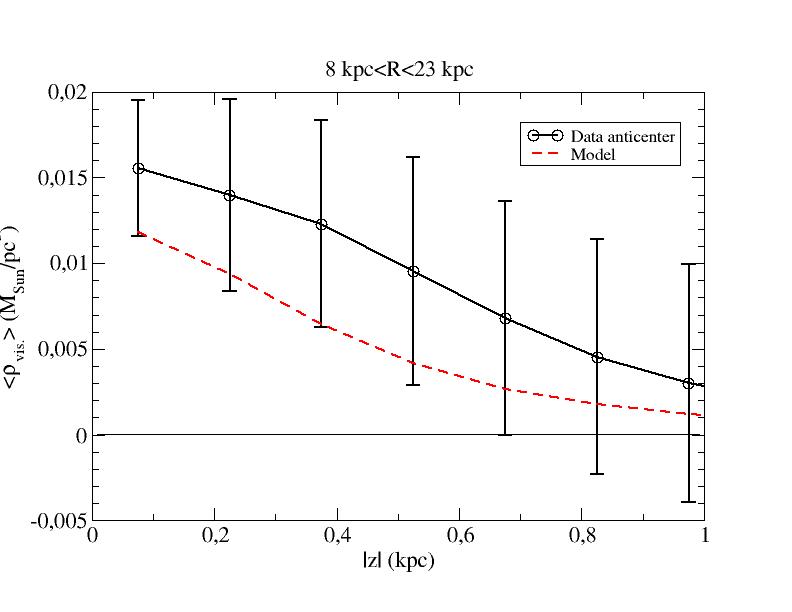}
\hspace{.2cm}
\includegraphics[width=8.5cm]{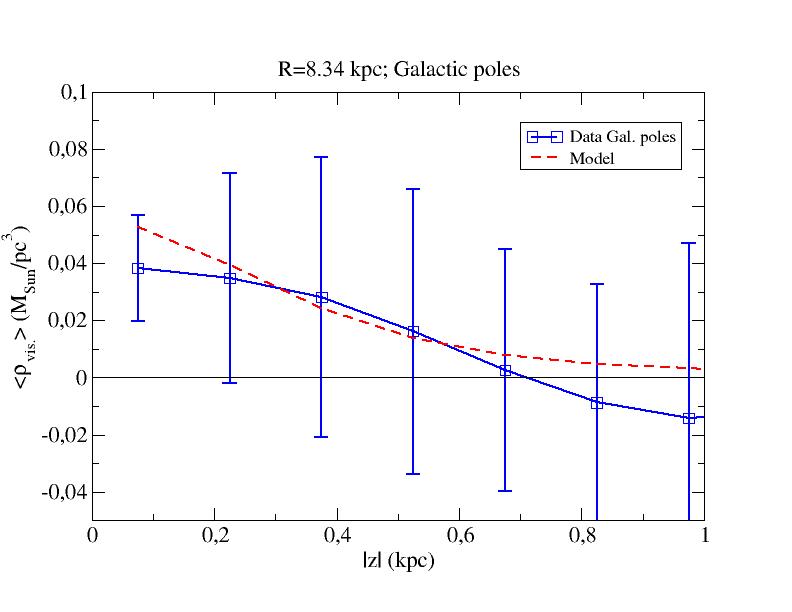}
\caption{Average visible matter density assuming MOND, in comparison with the model of visible matter. 
Top left: average within $|z|<0.15$ kpc,
as a function of $R$.
Top right: average within $|z|<1.05$ kpc, 
as a function of $R$. 
Middle left: average within 1.05$<|z|<2.10$ kpc,
as a function of $R$.
Middle right: average within 2.1$<|z|<2.7$ kpc, 
as a function of $R$.
Bottom left: average within 8 kpc$\le R\le 23$ kpc, as a function of $|z|$ within $|z|<1.05$ kpc in the anticenter region.
Bottom right: average density at $R=8$ kpc, as a function of $|z|$ within $|z|<1.05$ kpc in the Galactic pole regions.}
\label{Fig:roMOND}
\end{figure*}

Given that $\Sigma (R,z)=\Sigma _{\rm vis}(R,z)$ (there is no dark matter), one would expect that $\Sigma (R,z)$ in Fig.
\ref{Fig:sigmam} were the same as those given in Fig. \ref{Fig:sigmabar}. We get that.
 Fig. \ref{Fig:roMOND} illustrates this agreement. MOND works here.

Recently, \citet{Cha23} claimed that MOND is rejected at 5-$\sigma $ level in the fit of 
the rotation curve of Milky Way at $R$=17-23 kpc. 
However, other analyses \citep{Chr20,Wan23,Syl23} have shown that MOND may fit the rotation curve within the uncertainties, 
and as we have seen here,
the distribution of vertical velocities is also compatible with MOND and the visible (baryonic) matter knowledge.
Eq. (8) of \citet{Cha23} assumes that the rotation curve with MOND is the rotation curve with baryon matter multiplied by a factor of $1/\sqrt{\mu}$, an approximation that must be reconsidered when we try to fit data with small error bars. Indeed, a declining
rotation curve rather than a flat one is not incompatible with MOND \citep{Lop21}. In any case, rotation curves are
not the topic of this paper, but the dispersion of vertical velocities, which present no problem with MOND within the error bars.

\section{Discussion on possible systematic errors}
\label{.syst}

The results of our analyses can be summarized as follows: (i) the model of spherical dark matter halos of NFW and MOND model
compatible with the data; (ii) the model of the disky dark matter with density proportional to gas one (Bosma model) 
is totally excluded by the data [unless the scale height at $R<12$ kpc
is larger than 600 pc (2$\sigma $)]; and (iii) the total lack of dark matter
(there is only visible matter) within Newtonian gravity is not very far from the data (only a 
2.6$\sigma $ departure at $R=14$ kpc).

All of these results depend basically on four measurements or assumptions: the consideration that the
Jeans equations are applicable here, assuming equilibrium; the dispersion of
the vertical velocities; the average scale height of stellar distribution; and the vertical stellar density
distribution near the Galactic plane (sech$^2$ or exponential). We will discuss these
points in the next subsections.

\subsection{Applicability of the Jeans equations}

In this study, we have assumed that the stellar populations are in dynamical equilibrium, so that we can neglect the partial time derivatives in our theoretical framework, as 
is commonly done in these types of analyses \citep{Bov12,Mon12,Mon15,Che24}.
There may be some factors contributing to a time dependence, but the disequilibrium is negligible in the solar
neighborhood \citep[\S 5.4]{Che24} and the equilibrium necessary to apply the Jeans equations is suitable within
$R\lesssim 20$ kpc \citep{Chr20,Koo24} within errors lower than 10\%. 
Nonequilibrium may have an effect but as a secondary contribution that is not expected
to significantly change the qualitative results.
At $R<14$ kpc, where the most significant measurements are obtained, the uncertainty due to the application
of the Jeans equations is less than 2\% \citep{Koo24}.

\subsection{Reliability of the dispersion of vertical velocities}

Our results are very dependent on the measurement of $\sigma _{V_z}(R,z)$ in Eq. (\ref{fitsigmavz}).
As said previously, this dispersion of vertical velocities corresponds to the rms once
subtracted quadratically the measurement errors in radial velocities and proper motions
\citep[Sect. 4]{Lop19}, so it reflects the rms of the real distribution of velocities, and
not the dispersion due to errors in their measurements.
Independent calculations
\citep{Gai18,Nit21,Dri23,Che24} give
very similar results in the common range ($R<14$ kpc).

There is no assumption here, no theoretical priors, 
but just an analysis of Gaia data.
For the least distant sources ($R\lesssim 14$ kpc), the calculation of this dispersion is just
the result of the direct derivation of the distribution of velocities obtained with the component perpendicular to the plane of the proper motions (and a small component of radial velocities in off-plane regions) for the anticenter regions, and
the radial heliocentric velocities (and a small component of proper motions) for the Galactic pole regions. Assuming the measurements of Gaia DR3 are correct, the result cannot be changed. 
For the most distant sources ($R\gtrsim 14$ kpc), Lucy's method of deconvolution of parallax errors
\citep{Lop19} provides a correction to the statistical real distribution of distance.
In principle, all sources of errors in the calculation of distance errors with this
method have been taken into account, so there is no reason to doubt this result. Nonetheless,
even if we doubted the validity of the Lucy inversion technique, our conclusions would not
change because we see clearly the same conclusions for $R\lesssim 14$ kpc, where Lucy deconvolution has a negligible effect.

\subsection{The effect of the flare}

Another crucial element is the scale height and its variation with $R$ (flare).
It was derived from purely observational analysis of Gaia data stellar distribution, and it is not
dependent on theoretical assumptions. Nevertheless, we may speculate that
the scale heights are different for some reason. 
Instead, we will explore some variations of the flare model 
to see how it affects our results.

We will use the following modeling of scale heights of the stellar density instead of Eqs. (\ref{flare}) and (\ref{flarec}):
\begin{equation}
\label{flare2}
h_{z,{\rm thin}}(R)=0.20\,\exp [(R-R_\odot)/h_f] \ \ {\rm kpc}
\end{equation}\[
h_{z,{\rm thick}}(R)=0.65\,\exp [(R-R_\odot)/h_f] \ \ {\rm kpc}
,\]
where the normalizations are to be coincident at $R=R_\odot $ with Eqs. (\ref{flare}) and (\ref{flarec}), and
for simplicity we only have one free parameter, $h_f$ with the same flare for the thin and thick discs.
This type of model is indeed used in some other fits available in the 
literature \citep[e.g.,][]{Lop02,Li19}.

\begin{figure}
\vspace{1cm}
\includegraphics[width=8cm]{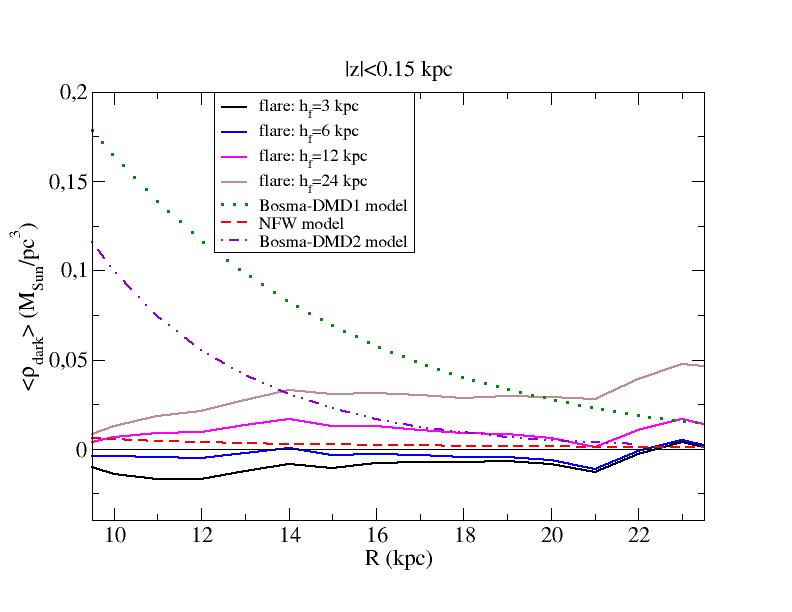}
%\vspace{.2cm}
%\includegraphics[width=8cm]{roMONDb2.jpg}
\caption{Average dark matter density in the anticenter regions with Newtonian gravity within $|z|<0.15$ kpc, assuming a scale height
with a flare given by Eq. (\ref{flare2}), where $h_f$ is a free parameter (lower h$_f$, stronger flare). 
For comparison, we show the predictions of the models of dark matter discussed in this paper.}
\label{Fig:flare2}
\end{figure}

In Fig. \ref{Fig:flare2} we can see the effect of the variation of $h_f$ for the density of
dark matter assuming Newtonian gravity. 
Within the range of $h_f$ between 3 and 24 kpc, the disagreement with the Bosma model is still very large, whereas the NFW is
more or less in agreement with any value of $h_f$.

\subsection{Vertical stellar density distribution near the Galactic plane}

\begin{figure}
\vspace{.2cm}
\includegraphics[width=8.5cm]{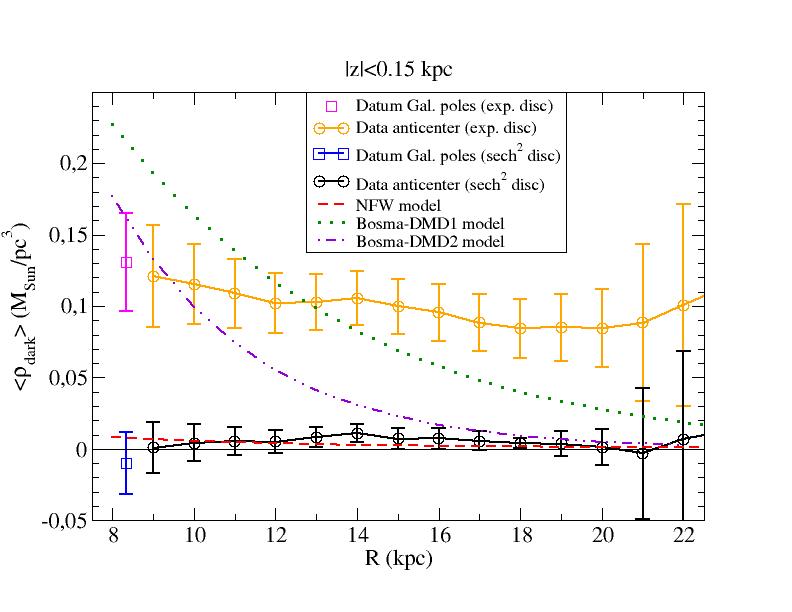}\\
\vspace{.2cm}
\includegraphics[width=8.5cm]{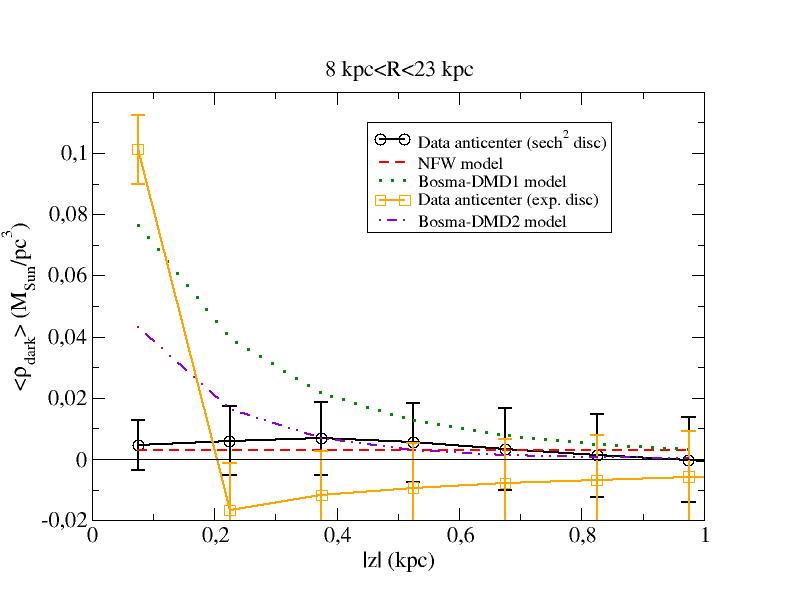}
\caption{Average dark matter density derived from Eq. (\ref{DM2}) assuming either
exponential or sech$^2$ (default) stellar vertical disk. Error bars include both the errors of the data
and the errors in the visible matter model. Top: average within $|z|<0.15$ kpc,
as a function of $R$. Bottom: average within 8 kpc$\le R\le 23$ kpc, 
as a function of $|z|$ within $|z|<1.05$ kpc in the anticenter region.
Comparison with theoretical models.}
\label{Fig:DMexp}
\end{figure}

We have assumed sech$^2$ functions in the vertical-direction stellar density, rather than an exponential one, in
order to keep the consistency of $\Sigma (z=0)=0$.
Nonetheless, we may explore the results of dark matter if we assume an exponential distribution (note that the normalization
is changed in Eq. \ref{baryonic} multiplying  $\rho_{M,*,\odot }$ and $\rho _{M,gas+dust,\odot }$ by a factor two with respect
to the sech$^2$ distribution in order to obtain the same surface density at $z=\infty $).
The dark matter density for this assumption is plotted in Fig. \ref{Fig:DMexp}.

The major difference with respect to sech$^2$ distribution 
is for $|z|<0.15$ kpc, where an exponential distribution would increase $\Sigma $
very significantly and consequently make $\rho _{\rm dark}$ larger. Nonetheless, the dependence on $R$ is
almost flat, which is still incompatible with the Bosma model (and also incompatible with the NFW model); and the density for
$|z|\gtrsim 0.2$ kpc fails to reproduce the Bosma model too.

The fact we get such a significant difference in the results of dark matter density 
between exponential and sech$^2$ distributions at $|z|<0.15$ kpc is precisely due to the differences
in both stellar distributions there. The high density of dark matter in the plane regions in comparison with
other values at higher $|z|$ makes no sense, and this reinforces the idea that an exponential stellar distribution should
not be used. In any case, it does not solve the problems for the Bosma model either.

\section{Further discussion and conclusions}
\label{.concl}

Our analysis applies the Jeans equations to the stellar velocity distribution of Gaia DR3 stars in the
anticenter and the Galactic poles to obtain the dynamical mass distribution near the plane regions.
With the use of the extended kinematic maps produced with a parallax error deconvolution
technique \citep{Lop19}, we are able to reach distances up to $R=22$ kpc.
Once we have the total mass density as a function of $R$, $z$ (we assume an axisymmetric disk),
we subtract the amount due to visible matter, and we obtain dark matter density.
Alternatively, substituting Newtonian gravity for MOND gravity, we obtain directly
the total density equal to the visible matter density.

This type of analysis has been carried out previously \citep[e.g.,][]{Mon12,Loe12,Vil22,Che24},
but not in a way as complete as we did here. Here, apart from using the most accurate velocity maps of Gaia DR3,
we carry out a careful analysis of the many parts of the Jeans equations in the vertical direction.
In particular, we realize how important it is to introduce a flare in the disk, especially at the largest $R$, or
the inclusion of a double thin+thick disk rather than a single disk, the dependence of the scale height
on the average age of the population, or more importantly, the stellar density
distribution near the plane. An accurate measurement of the vertical velocity dispersion as a function of $R$,
$z$ is also important, and different selection effects, statistical methods, or fitting laws 
may vary the numbers (see Sects. \ref{.extended}, \ref{.rms}). 
In the end, we cannot have infinitely precise data to carry out our calculation, but
we carry out a careful analysis of the effect of the most important sources of error bars.

A key question among the mentioned ones is whether the vertical distribution of stellar density follows an
exponential law or sech$^2$ law. Different observational works \citep[e.g.,][]{Dob20,Eve22,Chr22,Vie23} prefer one or the other solution. However, from a theoretical point of view, there is no possibility to choose among both: only one of
them is consistent with the Jeans equations for a steady equilibrium disk, a sech$^2$, whereas the vertical exponential
leads to the inconsistent result of $\Sigma (R,z=0)\ne 0$. With this in mind, we have used sech$^2$ distributions
for the visible (stars and gas+dust) matter, though in a few examples, we have checked the possible effects of changing to an exponential distribution, too.

Perhaps the most important result of the obtained surface density, $\Sigma (R,z)$, is that
its error bars are quite large if one takes into account all of the factors that contribute to the Jeans equations and
the modeling of the visible/baryonic matter. However, even with large error bars, we are able to set some modest
constraints in the distribution of dark matter (within Newtonian gravity) or the compatibility of MOND.
The results of our analyses can be summarized as follows: (i) the model of the spherical dark matter halos in the
NFW and MOND models are compatible with these data, though our data are not precise enough to corroborate the exact shape
of $\rho _{\rm dark}$; (ii) the model of the disky dark matter with density proportional to the gas density (Bosma model) 
is totally excluded by the data; iii) the total lack of dark matter
(there is only visible matter) within Newtonian gravity, is not very far from the data: 
only a 2.6$\sigma $ departure at $R=14$ kpc,
$|z|<1.05$ kpc; taking into account that we have analyzed hundreds of bins of $R$ and $z$, the probability of finding
one with a departure of 2.6$\sigma $ is likely.

\begin{enumerate}
\item
The first point is not a surprise and is in line with the results previously obtained by other researchers.
Nevertheless, we would like to remark that compatibility does not mean confirmation of a model. Further research is necessary
to be able to confirm the existence of these dark matter halos or modified gravity scenarios.

\item
The second point is not a surprise either, although it serves to remind us of a serious obstacle in the development
of a hypothesis that explains the rotation curve of spiral galaxies in terms of dark matter distributed in a disk rather than
in a halo \citep{Bos81,Pfe94,Fen15,McK15,Fer16,Kra16,Sip21,Syl23,Syl24,Syl24b}. Indeed, it has been well known for many decades  \citep{Kuz52,Kuz55} that there is no evidence for the presence of dark matter in the disk of the Galaxy (from observations in the
solar neighborhood). What we get here is that a disky dark matter predicts a dark matter density in in-plane regions ($|z|<0.15$ kpc) much higher than observed: we get
$\langle \rho _{\rm dark}\rangle (R,|z|<0.15)\lesssim 0.01$ M$_\odot $/pc$^3$ $\forall R$ between 8 and 12 kpc
(see Fig. \ref{Fig:DM}, top--left panel), whereas the disky dark matter between 0.05 and 0.2 M$_\odot $/pc$^3$ within that range of $R$ is
compatible with the rotation curves.
The only way to save the model of disky dark matter is decoupling from gas density and allowing a much larger
scale height [we would require a scale height at $R<12$ kpc
to be larger than 600 pc (2$\sigma $], which makes it closer to a spheroidal distribution, or 
substituting the sech$^2$ stellar density for an exponential profile, which may solve the problem in in-plane regions but however
is still inconsistent at larger $|z|$ values.

\citet{Syl24b} presents an opposite conclusion about the disky dark matter, using the same data. He claims
that a DMD2 model fits the data with $h_z\approx 100$ pc, of the order of the gas scale height, so it is compatible with Bosma hypothesis. However, this analysis by \citet{Syl24b} has a series of oversimplifications and wrong assumptions
that make his result invalid:
\begin{enumerate}
\item The hypothesis that the velocity distribution of dark matter particles and stars is the same one \citep[Eq. (12)]{Syl24b} is not appropriate since the particles of a putative non-baryonic dark matter would have different origins and formations. 
\item The use of Equations (21) and (22) of \citet{Syl24b} to calculate the average
scale height includes dark matter and gas density distributions, which is not correct; the scale height should be that
of the population of stars that are used to measure the dispersion of velocities (the stars), as stated in the Jeans equations (see the paragraph after Eq. (\ref{az}) in this paper).
\item There is a factor 2 in the Equations (28), (30), and (32) \citet{Syl24b} 
that is not correct. 
\item Sylos Labini uses a mass of the thick disk equal to 90\% of the mass of the  
thin disk, which is too high because 
the dominant disk should be the thin one; also, a scale length of 4.5 kpc for the thin disk is doubtful when
using optical sources of Gaia (different from the scale length with near infrared surveys).
\item The total $\Sigma $ [\citep[Fig. 11/top panel]{Syl24b}; note that $a_z=2\pi G\Sigma $] cannot be used to fit and test the DMD2 model because $\Sigma $ is dominated by the visible matter, and we need to subtract the visible matter separately to analyze 
the residuals of dark matter, as carried out here in Section \ref{.dark}. 
\item The fit of $h_z(R)$ for DMD2 model
given in \citet[Fig. 11/bottom panel]{Syl24b} avoids the points with $|z|<0.5$ kpc, which are precisely
the ones giving the major discrepancy of a low $h_z$ in DMD2 
(see Fig. \ref{Fig:sigma}, bottom-left panel in this paper).
\item The use of exponential vertical distribution instead of sech$^2$ presents some differences in the 
regime $|z|<0.5$ kpc (see Fig. \ref{Fig:DMexp} of this paper). This might be a reason to avoid this range,
but in any case, this range is necessary if we want to analyze a DMD2 model with most of its mass within it.
DMD2 gives a local dark matter density in the solar neighborhood equal to 
$\rho _{DMD2,\odot }=0.18$ M$_\odot$/pc$^3$, which is totally at odds with any other estimation, either in the present work or in many other analyses (none of the numbers
allows a dark matter density larger than $\sim 0.02$ M$_\odot$/pc$^3$; see references at point of 1 of 
Section \ref{.compDM}).
\end{enumerate}

\item
The third point is remarkable, too. We see that the observed values of $\Sigma $ derived from the dispersion of velocities are
consistent with a mass distribution only with visible matter and Newtonian gravity. 
Previous analyses presented overwhelming evidence for a significantly nonzero value of 
$\rho _{\rm dark} (R_\odot ,z=0)=0.01-0.02$ M$_\odot $/pc$^3$  \citep[e.g.,][]{Gar12,Cat12,Pat15,Xia16,deS21,Ou24},
and $\Sigma (R=R_\odot ,z=1-1.1 {\rm kpc})=50-80$ M$_\odot $/pc$^2$ \citep[Table 1]{Hor24}\citep{Nit21}, larger than
the visible matter surface density and implying the necessity of dark matter.
However, we have seen in this paper how sensitive the derivations of $\rho _{\rm dark}$ are based on kinematic data
on the different assumptions modeling the stellar density distribution, 
and dark matter can only be detected through the kinematical effects. 
We obtained $\Sigma (R=R_\odot ,z=0.9\ {\rm kpc})=43\pm 11$ M$_{\odot }$ pc$^{-2}$,
$\Sigma (R=R_\odot ,z=1.05\ {\rm kpc})=39\pm 18$ M$_{\odot }$ pc$^{-2}$ (within Newtonian gravity), 
lower than the most common values in the literature and compatible with expected surface density 
for visible matter alone [$\Sigma _{\rm vis.}(R=R_\odot ,z=0.9\ {\rm kpc})=42.8$ M$_{\odot }$ pc$^{-2}$; $\Sigma _{\rm vis.}(R=R_\odot ,z=1.05\ {\rm kpc})=43.7$ 
M$_{\odot }$ pc$^{-2}$ from Fig. \ref{Fig:sigmabar}], thus allowing zero dark matter.
These larger error bars are not due to worse data or a more awkward technique, but to a more strict modeling of the stellar distribution (see Sect. \ref{.errors}). We agree with \citet{Che24}
that the measured mass density is highly dependent on the assumptions when using the Jeans equations.
In general, the analysis of the Jeans equations
is (stellar density) model dependent, and a 3\% relative error in $\Sigma (R=R_\odot ,z=1.1\ {\rm kpc})$, as given in \citet{Nit21}
looks too optimistic, given that the uncertainties in the scale height and $\sigma _{V_z}$ are much larger than this.

In other words, we cannot reject the
hypothesis of a dark matter halo (with very low density near the plane), but we cannot prove it either. 
Given that the proof of the existence of (non-baryonic) dark matter on Galactic scales is still being discussed, and
there are many inconsistencies and tensions pending to be solved \citep[\S 3.4]{Lop22}, and with rotation curves in the Milky
Way allowing a large range of solutions even without dark matter (MOND or also within Newtonian gravity), 
our results make us think that the hypothesis of the existence of dark matter halos is less solid than we thought.
Therefore, still, even in the epoch of accurate measurements of stellar velocities with {\it Gaia} data, we must take the hypothesis of the existence of dark matter with a grain of salt, which is more supported by cosmological speculations than
by the observations of galaxies.

\end{enumerate}
%%%%%%%%%%%%%%%%%%%%%%%%%%%%%%%%%%%%%%%%%

\begin{acknowledgements}
We thank the referee for a thorough and very helpful review and detailed comments, which improved the quality of the paper.
Thanks are given to Francesco Sylos Labini for valuable comments, and the fittings 
in Section 3.3 and Figure 3. Thanks are given to Juan Betancort-Rijo and Roberto
Capuzzo-Dolcetta for discussions on the applicability of the Jeans equations.
This research was supported by the Chinese Academy of Sciences 
President’s International Fellowship Initiative grant No. 2023VMB0001
and grant No. PID2021-129031NB-I00 from the Spanish Ministry of Science (MICINN).
This work has made use of data from the European Space Agency (ESA) mission Gaia (https://www.cosmos.esa.int/gaia), 
processed by the Gaia Data Processing and Analysis Consortium (DPAC, https://www.cosmos.esa.int/web/gaia/dpac/consortium). 
Funding for the DPAC has been provided by national institutions, in particular, the institutions participating 
in the Gaia Multilateral Agreement.
\end{acknowledgements}

\section*{ORCID for authors}

M. L\'opez-Corredoira: 0000-0001-6128-6274

\appendix

\section{Jeans equations for a multicomponent distribution}
\label{.Jeansmulti}

Given a single stellar population component $i$ with a given density $\rho _i=\frac{N_i}{\cal{V}}$ 
($\cal{V}$ is volume of
each bin; $N_i$ its number of stars), average velocity distribution $V_{z,i}$,
the Jeans equations for the vertical direction (the reasoning would be similar for the radial direction),
applying Eq. (\ref{az}), neglecting the terms with $V_RV_z$ (see Sect. \ref{.rms}), is
\begin{equation}
\label{azi}
a_z(R,z)=-\frac{1}{N_i(R,z)}\frac{\partial [N_i(R,z)V_{z,i}^2(R,z)]}{\partial z}
\end{equation}

On average for all components,
\begin{equation}
\overline{ \rho }=\frac{\sum _i N_i}{\cal{V}}
,\end{equation}
\begin{equation}
\overline{V_{z}^2}=\frac{\sum _i N_iV_{z,i}^2}{\sum _i N_i}
\end{equation}
\begin{equation}
\overline{\rho V_{z}^2}=\frac{\sum _i N_iV_{z,i}^2}{\cal{V}}=\overline{\rho }\overline {V_{z}^2}
\end{equation}
Hence, using Eq. (\ref{azi}),
\begin{equation}
\frac{\partial [\overline{\rho V_{z}^2}(R,z)]}{\partial z}=
-\frac{a_z(R,z)}{\cal{V}}\sum _i N_i(R,z) =-a_z(R,z)\overline{\rho }
.\end{equation}
Reordering the terms, we have the Jeans equations for the multicomponent system:
\begin{equation}
a_z(R,z)=-\frac{1}{\overline{ \rho }}\frac{\partial [\overline{\rho } \overline{V_{z}^2}(R,z)]}{\partial z}
.\end{equation}

\section{Calculation of $\sigma _{V_z}$ for a distribution without high-velocity stars}
\label{.nonhighvz}
	
In a distribution of velocities assumed to be Gaussian but without the tails of high-velocity stars
($|V_z|>V_c$), the calculation of $\sigma _{V_z}$ is
\begin{equation}
\sigma _{V_z}=\sqrt{\sigma _{V_z}^2 {\rm erf} \left(\frac{V_c}{\sqrt{2}\sigma _{V_z}}\right)+
\frac{2}{\sqrt{2\pi} \sigma_{V_z}}\left[\int_{V_c}^\infty dx\,x^2 
\exp\left(-\frac{x^2}{2\sigma _{V_z}^2}\right)\right] }
,\end{equation}
which is solved iteratively for $\sigma _{V_z}$, with the first iteration equal to 
\begin{equation}
\sigma _{V_z,0}=\sqrt{ \langle V_Z^2\rangle -\langle V_Z\rangle ^2 }
.\end{equation}

For our practical case of $V_c=100$ km/s, $V_c$ is 2.5-5 times $\sigma _{V_z,0}$, so this correction
is very small, and $\sigma _{V_z}$ is very close to $\sigma _{V_z,0}$. We could even neglect this correction.
In any case, the above equation is exact for any cutoff of the Gaussian tail, and the calculation
of $\sigma _{V_z}$ would not be affected if we set a 
tighter constraint in the selection of non-high-velocity stars.

\end{document}